\documentclass[5p]{elsarticle}

\usepackage{lineno,hyperref}
\modulolinenumbers[5]

\usepackage{amsmath}
\usepackage{amssymb}

\newcommand{\intd}{\, \mathrm{d}}
\newcommand{\Id}{\mathrm{Id}}
\newcommand{\var}{\mathrm{var}}
\newcommand{\cov}{\mathrm{cov}}
\newcommand{\E}{\mathrm{E}}
\newcommand{\MSE}{\mathrm{MSE}}

\journal{Neural Networks}

\usepackage{xcolor}

\begin{document}
	
	\begin{frontmatter}
		
		\title{Performance boost of time-delay reservoir computing by non-resonant clock cycle}
		
		\author[1,2]{Florian Stelzer\corref{cor}}
		\cortext[cor]{Corresponding author}
		\ead{stelzer@math.tu-berlin.de}
		
		\author[3,4]{Andr\'e R\"ohm}
		\ead{aroehm@mailbox.tu-berlin.de}
		
		\author[3]{Kathy L\"udge}
		\ead{kathy.luedge@tu-berlin.de}
		
		\author[1]{Serhiy Yanchuk}
		\ead{yanchuk@math.tu-berlin.de}
		
		\address[1]{Institute of Mathematics, Technische Universit\"at Berlin, D-10623, Germany}
		\address[2]{Department of Mathematics, Humboldt-Universit\"at zu Berlin, D-12489, Germany}
		\address[3]{Institute of Theoretical Physics, Technische Universit\"at Berlin, D-10623, Germany}
		\address[4]{Instituto de F\'isica Interdisciplinar y Sistemas Complejos, IFISC (CSIC-UIB), Campus Universitat de les Illes Balears, E-07122 Palma de Mallorca, Spain}
		
		\begin{abstract}
			The time-delay-based reservoir computing setup has seen tremendous
			success in both experiment and simulation. It allows for the construction
			of large neuromorphic computing systems with only few components.
			However, until now the interplay of the different timescales has not
			been investigated thoroughly. In this manuscript, we investigate the
			effects of a mismatch between the time-delay and the clock cycle for
			a general model. Typically, these two time scales are considered to
			be equal. Here we show that the case of equal or resonant
			time-delay and clock cycle could be actively detrimental and leads
			to an increase of the approximation error of the reservoir. In particular,
			we can show that non-resonant ratios of these time scales have maximal
			memory capacities. We achieve this by translating the periodically
			driven delay-dynamical system into an equivalent network. Networks
			that originate from a system with resonant delay-times and clock cycles
			fail to utilize all of their degrees of freedom, which causes the
			degradation of their performance. 
		\end{abstract}
		
		
	\end{frontmatter}

\section{Introduction}

Reservoir computing is a machine learning method, which was introduced
independently by both Jaeger \cite{Jaeger2001} as a mathematical
framework and by Maass et al. \cite{Maass2002} from a biologically
inspired background. It fundamentally differs from many other machine
learning concepts and is particularly interesting due to its easy
integration into hardware, especially photonics \cite{SAN17a,BRU18a}.
With the help of the reservoir computing paradigm, the naturally
occurring computational power of almost any dynamical or physical
system can be exploited. It is particularly valuable for solving the
class of time-dependent problems, which is usually more difficult to address with
artificial neural network-based approaches. A time-dependent problem
requires to estimate a target signal $(y(t))_{t\in\mathbb{T}}$ which
depends non-trivially on an input signal $(u(t))_{t\in\mathbb{T}}$,
the set of times $\mathbb{T}$ may be continuous or discrete. This
class of problems contains, in particular, speech recognition or time
series prediction \cite{Jaeger2004,Verstraeten2005,Verstraeten2006},
and also has great promise for error correction in optical data transmission
\cite{ARG18}. Furthermore, reservoir computing can be used to study
fundamental properties of dynamical systems in a completely novel
way \cite{PAT18}, enabling new ways of characterizing physical systems.

The main idea of reservoir computing is simple, yet powerful: A dynamical
system, the \textit{reservoir}, is driven by an input $u(t)$. The
state of the reservoir is described by a variable $x(t)$, which can
be high- or even infinite-dimensional. A linear \textit{readout mapping}
$x(t)\mapsto\hat{y}(t)$ provides an output. While the parameters
of the reservoir itself remain fixed at all times, the coefficients
of the linear mapping $x(t)\mapsto\hat{y}(t)$ are subject to adaptation,
i.e. the readout can be trained.

Reservoir computing is a supervised machine learning method. An input
$u(t)$ and the corresponding target function $y(t)$ are given as
a training example. Then optimal output weights, i.e. coefficients
of the linear mapping $x(t)\mapsto\hat{y}(t)$, are determined such
that $\hat{y}(t)$ approximates the target $y(t)$. This is analogous
to a conventional artificial neural network where only the last layer
is trained. The goal of this procedure is to approximate the mapping
$u(t)\mapsto y(t)$ via $u(t)\mapsto x(t)\mapsto\hat{y}(t)$ such
that it not only reproduces the target function for the given training
input but also provides meaningful results for other, in certain sense
similar, inputs. From a nonlinear dynamics perspective, the readout
mapping $x(t)\mapsto\hat{y}(t)$ is a linear combination of different
degrees of freedom of the system.

The reservoir must fulfill several criteria to exhibit good computational
properties: First, it must be able to carry information of multiple
past input states, i.e. have memory. References \cite{Jaeger2002}
and \cite{Lymburn2019} study how a reservoir can store information
about past inputs. Second, the system should contain a non-linearity
to allow for non-trivial data processing. Jaeger \cite{Jaeger2001}
proposed to use a recurrent neural network with random connections
as reservoir. In this case the reservoir $x$ has a high-dimensional
state space by design, as the dimension equals the number of nodes.
Such systems are in general able to store a large amount of information.
Moreover, the recurrent structure of the network ensures that information
about past input states remains for a number of time steps and fades
slowly. Jaeger compared the presence of past inputs in the state of
$x$ to echoes. For this reason he called his proposed reservoir system
\textit{echo state network}.

In recent years, the field of reservoir computing has profited from
experimental approaches that use a continuous time-delay dynamical
system as reservoir \cite{Appeltant2011}. While hybrid network-delay
systems have also been proposed \cite{ROE18a}, typically, only a
single dynamical nonlinear system is employed and connected to a long
delay loop; i.e. as opposed to a network-based approach only a single
active element is needed for a delay-based reservoir. Here, the complexity
is induced by the inherent large phase space dimension of the dynamical
system with time-delay \cite{Erneux2009,Erneux2017,Yanchuk2017}.
The main advantage of using a delay system over Jaeger's echo state
approach \cite{Jaeger2001} is that one can physically implement
the delay system with analogue hardware at relatively low costs\textemdash either
electronically \cite{Appeltant2011} or even optically \cite{Brunner2013,Larger2012}.

Two main time scales exist in such delay-based reservoir systems:
the delay-time $\tau$ given by the physical length of the feedback
line, and a clock cycle $\tau'$ given by the input speed. In this
paper, we show that the non-trivial cases of mismatched delay-time
and clock cycle possess better reservoir computing properties. We
explain this by studying the corresponding equivalent network, where
we can show that non-resonant ratios of $\tau'$ to $\tau$ have maximal
memory capacities.

The paper is organized as follows: In section~\ref{sec:tdr} the
general model of a reservoir computer based on a single delay-differential
system is introduced. We refer to this method as \textit{time-delay
reservoir computing} (TDRC). Section~\ref{sec:numerical_observations}
shows numerical simulations and the effect of mismatching clock cycles
and delay-times. Section~\ref{sec:network} derives a representation
of the TDRC system with mismatching clock cycle and delay-time as
an equivalent echo state network. Section~\ref{sec:mc} presents
a direct calculation of the memory capacity. Section~\ref{sec:analytics}
derives a semi-analytic explanation for the observed decreased memory
capacity for resonant $\tau'$ to $\tau$ ratios and provides an intuitive
interpretation. All results are summarized in section~\ref{sec:summary}.

\section{Time-delay reservoir computing}

\label{sec:tdr}

In this section we describe the reservoir computing system based on
a delay equation. Its choice is inspired by the publications \cite{Appeltant2011,Brunner2013},
where it was experimentally implemented using analogue hardware. In
comparison to the general reservoir computing scheme $u(t)\mapsto x(t)\mapsto\hat{y}(t)$
described above, an additional `preprocessing' step is added to transforms
the input $u$ in an appropriate way before being sent to the reservoir.
This is particularly necessary when the input $u$ is discrete and
the reservoir $x$ is time-continuous. In the following, we describe
the resulting chain of transformations 
\begin{equation}
u\stackrel{\mathrm{(I)}}{\mapsto}J(t)\stackrel{\mathrm{(II)}}{\mapsto}x(t)\stackrel{\mathrm{(III)}}{\mapsto}\hat{y}(t)\label{eq:rcs}
\end{equation}
in detail.

\subsection{Step \textup{(I):} preprocessing of the input}

Since the reservoir is implemented with the physical experiment in
mind (e.g. semiconductor laser), its state variable $x(t)$ is time-continuous.
However, the input data is discrete in typical applications of TDRC
\cite{Appeltant2011,ROE18a,Brunner2013}. For this reason the preprocessing
function $u\mapsto J(t)$ translates the discrete input $u$ into
a continuous function $J(t)$.

We consider a discrete input sequence $(u(k))_{k\in\mathbb{N}_{0}}$,
where $u(k)\in\mathbb{R}$ is one-dimensional, however, the method
can be extended to multi-dimensional inputs. The important parameters
that define the preprocessing are the clock cycle $\tau'>0$, number
of virtual nodes $N\in\mathbb{N}$ and the resulting time per virtual
node $\theta:=\tau'/N$. In Ref.~\cite{Appeltant2011} the parameter
value for the clock cycle was chosen $\tau'=\tau$ and the optimal
values for $\theta$ are shown to be  in the interval $[0.1,1]$.
Therefore, we will consider the parameter $\tau'$ of the order of
the delay $\tau$ and $\theta$ within the interval $[0.1,1]$ or
even larger. In fact the exact value of $\theta$, or correspondingly,
$N$, does not qualitatively influence the phenomenon of the error 
increase by rational $\tau'/\tau$ (as we show in the appendix). 

\begin{figure}
\centering \includegraphics[width=255pt]{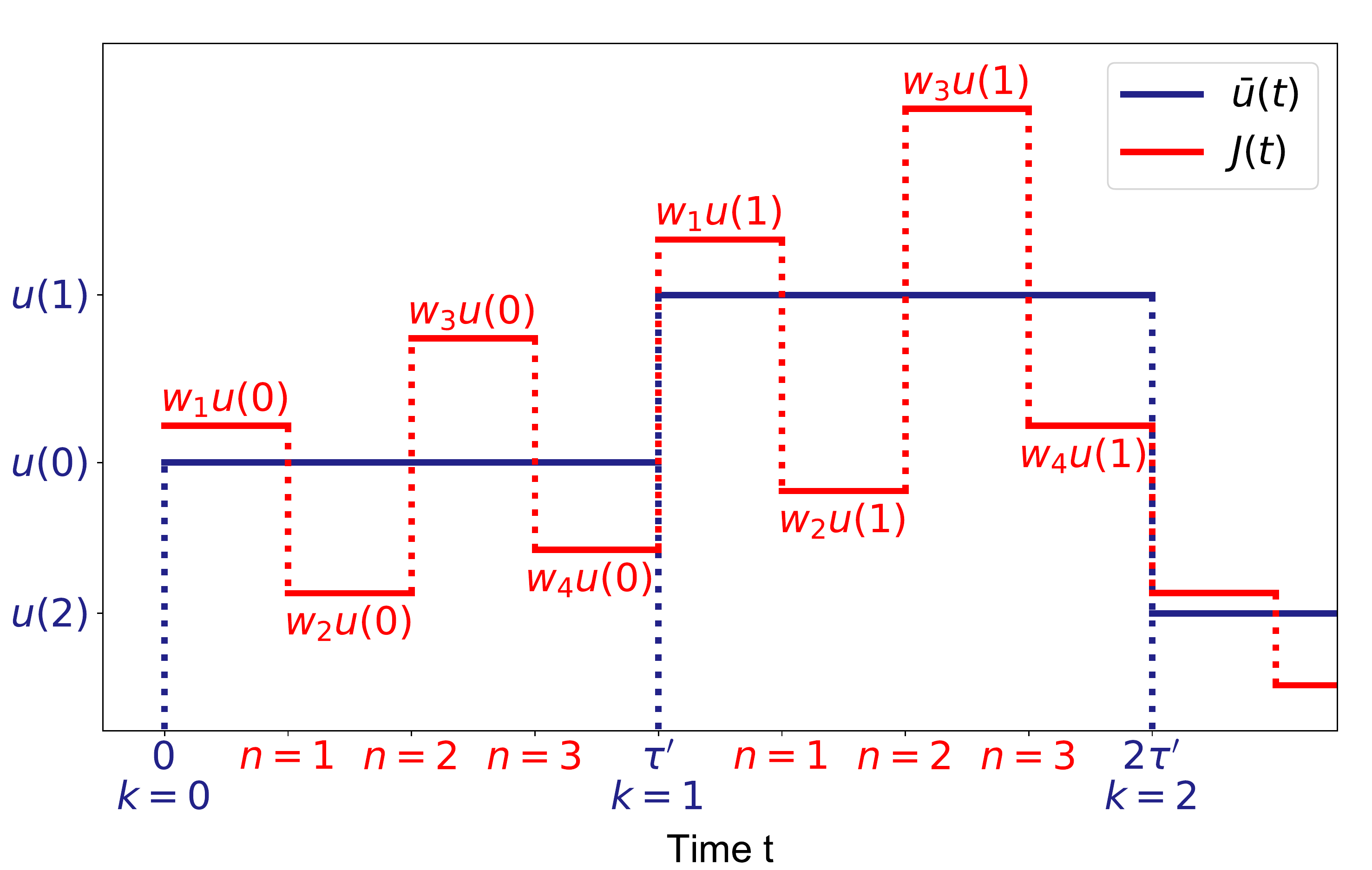}\caption{\label{fig:preprocessing} Schematic representation of the preprocessing
step: the discrete input sequence $u(0),\,u(1),\,\dots$ defines the
function $\bar{u}(t)$ (blue), which is multiplied by a $\tau'$-periodic
mask function $\mathcal{M}(t)$ to obtain the preprocessed input $J(t)$
(red). Here the length of the mask vector is $N=4$. The resulting
function $J(t)$ enters the reservoir equation~\eqref{eq:delay_system}. }
\end{figure}

First, a function $\bar{u}(t)$ is defined as step function 
\begin{align}
\bar{u}(t):=u(k),\quad t\in[k\tau',(k+1)\tau'),\ k\in\mathbb{N}_{0}
\end{align}
with step length $\tau'$. Using the indicator function 
\begin{equation}
\Pi_{M}(t)=\begin{cases}
1, & t\in M,\\
0, & t\not\in M,
\end{cases}
\end{equation}
the definition of $\bar{u}$ can be equivalently written as 
\begin{equation}
\bar{u}(t)=\sum_{k\in\mathbb{N}_{0}}u(k)\Pi_{[k\tau',(k+1)\tau')}(t).
\end{equation}
Secondly, $\bar{u}(t)$ is multiplied by the $\tau'$-periodic mask
\begin{equation}
\mathcal{M}(t)=\sum_{n=1}^{N}w_{n}\Pi_{[(n-1)\theta,n\theta)}(t\,\text{mod}\,\tau'),
\end{equation}
which is piecewise constant with step length $\theta=\tau'/N$ and
values $w_{n}$. Multiple options for the choice of a mask function
are compared in reference~\cite{Kuriki2018}. The final preprocessed
input signal $J(t)$ is 
\begin{align}
\begin{split}J(t) & :=\mathcal{M}\left(t\right)\bar{u}(t)\\
 & \hphantom{:}=\sum_{\substack{k\in\mathbb{N}_{0}\\
1\le n\le N
}
}w_{n}u(k)\Pi_{[k\tau'+(n-1)\theta,k\tau'+n\theta)}(t).\label{eq:J}
\end{split}
\end{align}
It is a piecewise constant function with values 
\begin{equation}
J_{k,n}:=w_{n}u(k)\label{eq:J_kn}
\end{equation}
on the intervals $[k\tau'+(n-1)\theta,k\tau'+n\theta)$. The details
of the preprocessing are illustrated in figure~\ref{fig:preprocessing}.

For further analysis, it is convenient to denote the `mask'-vector
$W^{\text{mask}}$ and the input vector $J_{k}$ as follows: 
\begin{equation}
W^{\text{mask}}:=(w_{1},\dots,w_{N})^{T},\quad J_{k}=W^{\text{mask}}u(k).\label{eq:Wmask-Jk}
\end{equation}

\subsection{Step \textup{(II):} reservoir}

Inspired by the previous works \cite{Appeltant2011}, we study the
reservoir given by the delay-diffential equation 
\begin{align}
\frac{dx}{dt}(t)=-x(t)+f[x(t-\tau)+\gamma J(t)],\label{eq:delay_system}
\end{align}
where $\tau>0$ is the delay-time, $\gamma>0$ is the input strength,
$f\colon\mathbb{R}\to\mathbb{R}$ the activation function, and $J(t)$
the preprocessed input function. For a given preprocessed input $J(t)$,
the reservoir variable $x(t)$ is computed by solving the delay differential
equation \eqref{eq:delay_system} with an initial history function
$x_{0}(s),$ $s\in[-\tau,0]$. In order for the reservoir to be consistent,
the reservoir state $x(t)$ for sufficiently large $t>0$ should depend
only on the input $J$ and be independent on
the initial state $x_{0}(s)$. In other words, identical reservoirs
which are driven by the same input but have different initial states
will approximate each other asymptotically. In the literature about
reservoir computing this property is often referred to as \textit{echo
state property} or \textit{fading memory property} \cite{Jaeger2001}.
In the literature about dynamical systems, the phenomenon is called
\textit{generalized synchronization} \cite{Rulkov1995,Pikovsky2001}
or \textit{asymptotic stability} \cite{Hale1993,Smith2010}. 

\subsection{Step \textup{(III):} readout}

The continuous reservoir variable $x(t)$ needs to be discretized
for the output. For this, the dynamical system is read out every $\theta$
time units. Because every small time window $[k\tau'+(n-1)\theta,k\tau'+n\theta)$
is fed with its own input $J_{k,n}$, these time windows are often
seen as `virtual nodes', and the whole delay system as a `virtual
network' \cite{Appeltant2011,Schumacher2013}. We discretize the
reservoir variable correspondingly: 
\begin{align}
X(k):=\begin{pmatrix}X_{1}(k)\\
\vdots\\
X_{N}(k)
\end{pmatrix}:=\begin{pmatrix}x((k-1)\tau'+\theta)\\
x((k-1)\tau'+2\theta)\\
\vdots\\
x((k-1)\tau'+N\theta)
\end{pmatrix}.\label{eq:vector_X}
\end{align}
In fact, $X(k)$ is the vector containing the $N$-point discretization
of the variable $x(t)$ on the interval $((k-1)\tau',k\tau']$. The
output $\hat{y}=(\hat{y}(k))_{k\in\mathbb{N}_{0}}$ of the machine
learning system is defined as 
\begin{align}
\hat{y}(k)=W^{\mathrm{out}}X(k)+c,\label{eq:readout}
\end{align}
where $W^{\mathrm{out}}$ is an $N$-dimensional row vector and $c\in\mathbb{R}$
is a scalar bias. The output weight variables $W^{\mathrm{out}}$
and $c$ are to be adjusted in the training process and are chosen
by linear regression for reservoir computers \cite{Jaeger2001}.

\section{Effect of the mismatch between delay and clock cycle times\label{sec:numerical_observations}}

When TDRC was first introduced, the clock cycle $\tau'$ for the preprocessing
of the input mask was chosen to be equal to the delay $\tau$ \cite{Appeltant2011,Brunner2013}.
In this case one can easily find an `equivalent network' which is
a discrete approximation of the reservoir system. See the suplementery
material of \cite{Appeltant2011} or reference~\cite{SCH13l} for
an example. However, recent numerical observations show, that the
performance may be improved if one sets $\tau'\neq\tau$. The earliest
example of this can be arguably found in reference~\cite{PAQ12}.

\begin{figure}
\centering \includegraphics[width=255pt]{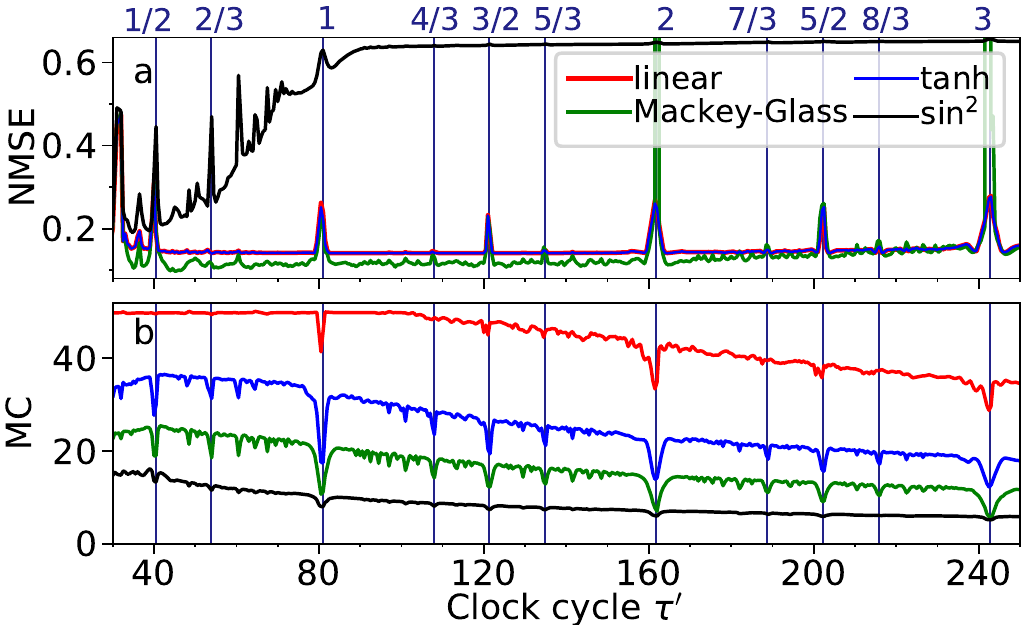}
\caption{Normalized mean square error (NMSE) of four different
TDRCs given in Eq.~\eqref{eq:delay_system} with nonlinearities of Eqs.~\eqref{eq:f_linear}-\eqref{eq:f_sin} for the NARMA-10 task (top) and the total memory capacity $\mathrm{MC}_{\mathrm{tot}}$
(bottom). Clearly visible are error peaks for values of $\tau'$ that
are close to integer multiples of $\tau$ and low-order resonances.
The presence of the peaks does not depend on the type of the activation
function. Vertical lines denote the resonant values of $\tau$ to $\tau'$ as indicated on the top axis.
The parameters for the simulations are listed in table \ref{table:parameters}.}
\label{fig:system_error} 
\end{figure}

\begin{figure}
\centering \includegraphics[width=255pt]{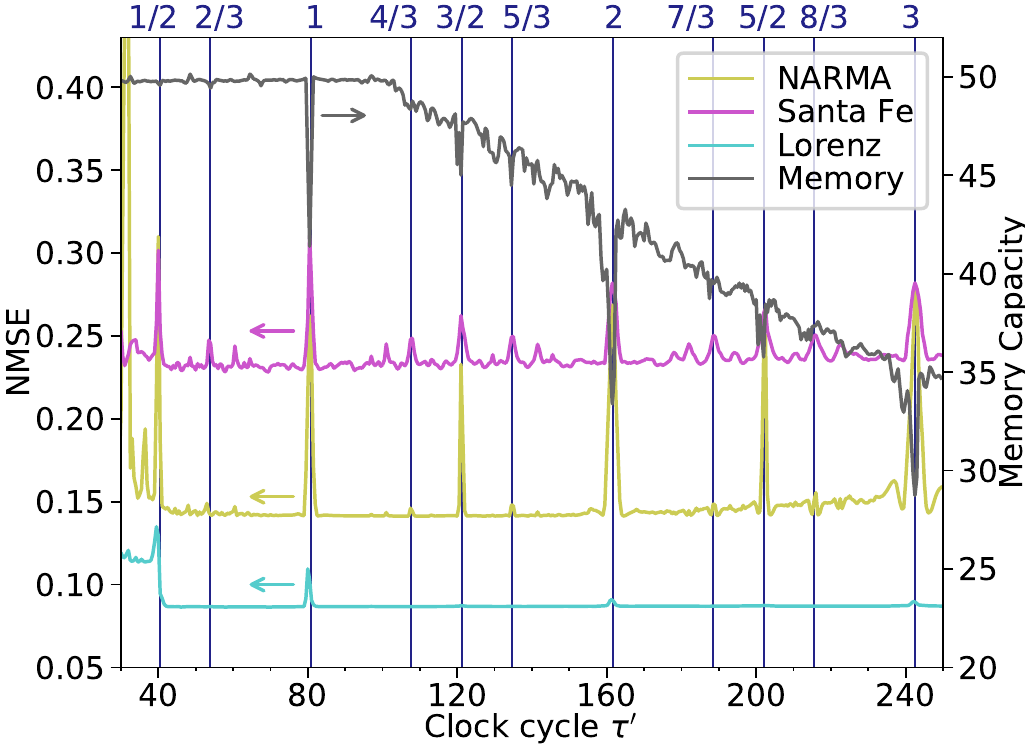}
\caption{Normalized mean square error (NMSE) for the NARMA-10
task, the Santa Fe time-series task and the Lorenz task (left axis)
and the total memory capacity $\mathrm{MC}_{\mathrm{tot}}$ (right
axis) of the TDRC of Eq.~\eqref{eq:delay_system} with linear activation function Eq.~\eqref{eq:f_linear}. For all tasks error
peaks (resp. memory capacity drops) are visible for values of $\tau'$
that are close to integer multiples of $\tau$ and low-order resonances.
Vertical lines denote the resonant values of $\tau$ to $\tau'$ as indicated on the top axis. The
parameters for the simulations are listed in table \ref{table:parameters}.}
\label{fig:tasks_error} 
\end{figure}

We use the NARMA-10 task \cite{Atiya2000}, the memory capacity (MC)
\cite{Jaeger2002}, the time series prediction task for the chaotic Lorenz system
and the Santa Fe time-series prediction task \cite{Weigend1993} to measure the performance of
a simple TDRC to illustrate the role of the clock cycle $\tau'$.
These are typical benchmark tasks and we refer to Appendix~\ref{subsec:narma}-\ref{subsec:santafe} for
a detailed explanation.

Moreover, we use four different functions for the activation function
$f$ in equation~\eqref{eq:delay_system}, 
\begin{align}
 & \text{linear:} & f(x) & =\alpha x,\label{eq:f_linear}\\
 & \text{Mackey-Glass:} & f(x) & =\alpha\frac{x}{1+x^{p}},\\
 & \text{hyperbolic tangent:} & f(x) & =\alpha\tanh(x),\\
 & \text{quadratic sin:} & f(x) & =\alpha\sin^{2}(x+\varphi),\label{eq:f_sin}
\end{align}
where we use the parameter values $p=1$ and $\varphi=0.5$. The
activation factor is chosen $\alpha=0.9$ for all cases. In fact,
the precise choice of the parameter $\alpha$ does not influence qualitatively
the phenomenon which we consider, see Appendix \ref{subsec:parameters} for more
details. The other parameters are set to $\tau=80$, $N=50$, and
$\gamma=0.02$. For our tests we choose $100$ different input weight
vectors $W^{\mathrm{mask}}$ with independently $\mathcal{U}(-1,1)$-distributed
entries $w_{n}$. For each input weight vector we train the system
with $50000$ training time steps and a Tikhonov regularization with
parameter $\beta=10^{-8}$. The training starts after an initial
period of $200000$ inputs which is necessary to ensure that the
reservoirs initial state does not influence the results. Only then do we record the system state for the next $20000$ inputs, with which we facilitate the training. The
only exception are the simulations for the Santa Fe time-series prediction
task. Since the Santa Fe dataset contains only about $9000$ data points,
we choose aperiod of $4000$ initial inputs, $4000$ training
steps and $1000$ test steps. All parameters for the simulations
are summarized in table~\ref{table:parameters}. The memory capacity is evaluated 
up to length $300$. Values that are close to $0$ are not included in the sum. We estimate
this overfitting threshold dynamically during the run, by using uncorrelated random target variables.

\begin{table}
	\centering
	\begin{tabular}{ ll }
		\hline
		\multicolumn{2}{c}{\underline{system parameters}} \\
		delay-time $\tau$          & $80$ \\
		input scaling $\gamma$     & $0.02$ \\
		number of nodes $N$        & $50$ \\
		activation factor $\alpha$ & $0.9$   \\
		range of clock cycle $\tau'$  & $[30,235]$  \\
		Mackey-Glass exponent $p$  & $1$   \\
		phase shift $\varphi$ in activation function \eqref{eq:f_sin}  & $0.5$ \\
		\hline
		\multicolumn{2}{c}{\underline{simulation parameters}} \\
		regularization parameter $\beta$ & $10^{-8}$ \\
		number of runs (with random mask) & $100$ \\
		number of initial steps per run: &  \\
		$\quad$ $\circ$ for Santa Fe task & $4000$ \\
		$\quad$ $\circ$ for all other task & $200000$ \\
		number of training steps per run: & \\
		$\quad$ $\circ$ for Santa Fe task & $4000$ \\
		$\quad$ $\circ$ for all other task & $50000$ \\ 
		number of test steps per run: & \\
		$\quad$ $\circ$ for Santa Fe task & $1000$ \\
		$\quad$ $\circ$ for all other task & $20000$ \\
		time step of $4$th order Runge-Kutta & 0.01\\
		\hline
	\end{tabular}
	\caption{Parameters for the numerical computation of the NMSE for the NARMA-10, Lorenz and Santa Fe task and the memory capacity of the system.}
	\label{table:parameters}
\end{table}

The top panel of figure~\ref{fig:system_error} shows the results
of simulations for the NARMA-10 task for the four different activation
functions \eqref{eq:f_linear}-\eqref{eq:f_sin}. The figure shows
clear peaks of the error for certain values of the clock cycle $\tau'$.
These peaks are located close to low-order resonances with the delay-time
$\tau=80$ and fulfill the relation $a\tau\approx b\tau'$ for small
$a,b\in\mathbb{N}$. In fact, the peaks are located slightly above
the resonant $\tau'$ values. Note, that $tanh$ and the linear function have very similar values.
The lower panel of figure~\ref{fig:system_error} depicts the memory capacity and
reveals at least part of the reason for this: The total memory capacity
for these resonant clock cycles decreases dramatically. 

To explore if the observed effect is related to the choice of task, we plot the performance for four
different tasks in Figure~\ref{fig:tasks_error}. In each case the linear activation \eqref{eq:f_linear}
was used. One can see that the error peaks (resp. performance drops)
are visible in all tested tasks.

The rest of this paper is devoted to the explanation of this phenomenon.
Thereby we focus on the linear activation \eqref{eq:f_linear}. Lacking a nonlinearity, this
is not the optimal choice for the TDRC, the resonance effects
that we are interested in seem to be general and independent of $f$
as shown in figure~\ref{fig:system_error}, where the performance for the activation functions~\eqref{eq:f_linear}-\eqref{eq:f_sin}
is computed. Furthermore, this simplification will allow us to deduce analytical results. In particular, we will be able to show the reason for the drop of the linear memory capacity explicitly. 

\section{Approximation by a network}

\label{sec:network}

In order to explain the degradation of the memory capacity for the
resonant clock cycles, we present the time delay-system reservoir as an
equivalent network. Similar procedure was done in \cite{Appeltant2011}
for a TDRC with 
$\tau'=\tau$ and arbitrary activation function. Here we present a
derivation of an equivalent network for the case $\tau'\neq\tau$,
where the results of \cite{Appeltant2011} cannot be applied. However,
we have to restrict it to a linear activation function in order to obtain
an explicit network representation.

An alternative way to describe
the dependence of $X(k+1)$ on $X(k)$ and the input for the case $\tau' \neq \tau$ is presented
in \cite{Larger2017}, where the authors use an integral formula instead of a network formulation. This integral formula includes the case of a nonlinear activation function. For our explanation of the observed memory capacity drops, the network representation presented in this section is, however, indispensable.

Since a detailed derivation is technical, we move it to Appendix \ref{subsec:network}
and present the main results in this section.

As follows from Appendix \ref{subsec:network}, the TDRC dynamics
can be approximated by the discrete mapping 
\begin{align}
\begin{split}\tilde{X}(k+1)=B\tilde{X}(k)+(1-e^{-\theta})F\left[A_{q}\tilde{X}(k+1-\ell)\right.\\
\left.+A_{-(N-q)}\tilde{X}(k-\ell)+\gamma J_{k}\right],
\end{split}
\label{eq:general_network}
\end{align}
where 
\begin{align}
\tilde{X}(k):=A_{0}^{-1}X(k)\label{eq:X_tilde_def}
\end{align}
and $X(k)$ is the discretized vector of the reservoir defined in
equation~\eqref{eq:vector_X}. Let us define and explain further
notations used in the mapping~\eqref{eq:general_network}. The matrix
\begin{align}
A_{0}:=\begin{pmatrix}1 & 0 & \dots & 0\\
e^{-\theta} & 1 & \ddots & \vdots\\
\vdots & \ddots & \ddots & 0\\
e^{-(N-1)\theta} & \dots & e^{-\theta} & 1
\end{pmatrix}
\end{align}
is the classical coupling matrix of an equivalent network for TDRC
with $\tau'=\tau$ \cite{Appeltant2011}. Moreover, 
\begin{align}
\begin{split}\ell & :=\left\lfloor \frac{m}{N}\right\rfloor ,\quad q:=m\text{ mod }N,\\
m & :=\left\lceil \frac{\tau}{\theta}\right\rceil =\left\lceil \frac{\tau}{\tau'}N\right\rceil ,\quad\theta=\frac{\tau'}{N},
\end{split}
\label{eq:mql_def}
\end{align}
where $\lfloor\cdot\rfloor$ and $\lceil\cdot\rceil$ denote the floor
and the ceiling function, which we need to employ to allow delay-times
$\tau$ that are not an integer multiple of the time per virtual node $\theta$. These quantities
can be interpreted as follows: $m$ is the number of virtual nodes
that are needed to cover a $\tau$-interval, $q$ is a measure of
the misalignment between $\tau$ and $\tau'$, and $\ell$ is roughly
the ratio between the delay-time $\tau$ and the clock cycle $\tau'$.
For $\ell=0$, the delay $\tau$ is shorter than the clock cycle $\tau'$,
and it is similar to or larger than the clock cycle for $\ell\geq1$.
The matrices $A_{q}$ and $A_{-(N-q)}$ are shifted versions of the
matrix $A_{0}$. They are defined as follows: 
\begin{equation}
A_{q}:=\begin{pmatrix}0 & \cdots & \cdots & \cdots & \cdots & \cdots & 0\\
\vdots &  &  &  &  &  & \vdots\\
0 &  &  &  &  &  & \vdots\\
1 & 0 &  &  &  &  & \vdots\\
e^{-\theta} & 1 & \ddots &  &  &  & \vdots\\
\vdots & \ddots & \ddots & 0 &  &  & \vdots\\
e^{-(N-1-q)\theta} & \cdots & e^{-\theta} & 1 & 0 & \cdots & 0
\end{pmatrix}
\end{equation}
is obtained by a downwards shift of $A_{0}$ by $q$ rows and 
\begin{multline}
A_{-(N-q)}:=
\\
\begin{pmatrix}e^{-(N-q)\theta} & \cdots & e^{-2\theta} & e^{-\theta} & 1 & 0 & \cdots & 0\\
\vdots &  &  & \ddots & \ddots & \ddots & \ddots & \vdots\\
\vdots &  &  &  & \ddots & \ddots & \ddots & 0\\
e^{-(N-1)\theta} & \cdots & \cdots & \cdots & \cdots & e^{-2\theta} & e^{-\theta} & 1\\
0 & \cdots & \cdots & \cdots & \cdots & 0 & 0 & 0\\
\vdots &  &  &  &  &  &  & \vdots\\
0 & \cdots & \cdots & \cdots & \cdots & \cdots & \cdots & 0
\end{pmatrix}
\end{multline}
is obtained by the upwards shift of $A_{0}$ by $N-q$ rows. Furthermore
\begin{align}
B:=\begin{pmatrix}e^{-N\theta} & \dots & e^{-\theta}\\
0 & \dots & 0\\
\vdots &  & \vdots\\
0 & \dots & 0
\end{pmatrix},
\end{align}
\begin{align}
F\begin{pmatrix}x_{1}\\
\vdots\\
x_{N}
\end{pmatrix}:=\begin{pmatrix}f(x_{1})\\
\vdots\\
f(x_{N})
\end{pmatrix},
\end{align}
and $J_{k}$ is defined as the input vector~\eqref{eq:Wmask-Jk}.

The mapping~\eqref{eq:general_network} generalizes previous results
from \cite{Appeltant2011,Schumacher2013}. If the clock cycle $\tau'$
satisfies $\tau'\in[\tau,\tau+\theta)$, the description coincides
with the classical case $\tau'=\tau$ and the approximate equation~\eqref{eq:general_network}
yields the same mapping 
\begin{align}
\tilde{X}(k+1)=B\tilde{X}(k)+(1-e^{-\theta})F[A_{0}\tilde{X}(k)+\gamma J_{k}]
\end{align}
as presented in \cite{Appeltant2011} because then $\ell=1$, $q=0$,
and $A_{-(N-q)}=0$.

Analytical approaches for the nonlinear system~\eqref{eq:general_network}
are challenging. To simplify, we will study the effect of different
clock cycles $\tau'$ with the help of a linear activation function
$f(x)=\alpha x$, where $\alpha$ is a scalar. Then equation~\eqref{eq:general_network}
can be written as 
\begin{align}
\begin{split}\tilde{X}(k+1) & =B\tilde{X}(k)+\nu\alpha[A_{q}\tilde{X}(k+1-\ell)\\
 & \hphantom{{}=}+A_{-(N-q)}\tilde{X}(k-\ell)+\gamma W^{\mathrm{mask}}u(k)]
\end{split}
\label{eq:linear_network_tau}
\end{align}
by plugging in $W^{\mathrm{mask}}u(k)$ for $J_{k}$ and by writing
$\nu:=1-e^{-\theta}$ for the sake of shortness.

System~\eqref{eq:general_network} possesses the following properties:
in the case $\tau'\geq\tau+\theta$, we have $\ell=0$, and hence,
equation~\eqref{eq:general_network} is in general an implicit map.
This is the physical case of a delay shorter than the clock cycle,
which means that the feedback from some of the virtual nodes will
act on other virtual nodes within the same cycle. However, for the
linear activation function in equation~\eqref{eq:linear_network_tau}
and by \eqref{eq:X_tilde_def} we obtain for the case $\ell=0$ the
explicit map 
\begin{align}
X(k+1)=AX(k)+W^{\mathrm{in}}u(k),\label{eq:linear_network}
\end{align}
where 
\begin{align}
A:=A_{0}(\Id-\nu\alpha A_{q})^{-1}(B+\nu\alpha A_{-(N-q)})A_{0}^{-1}
\end{align}
is a matrix that describes the coupling and local dynamics of the
virtual network and 
\begin{align}
W^{\mathrm{in}}:=\nu\alpha\gamma A_{0}(\Id-\nu\alpha A_{q})^{-1}W^{\mathrm{mask}}
\end{align}
is the generalized input matrix. 

Equation~\eqref{eq:linear_network} is the main result of this section.
It shows that the TDRC system can be modeled by an equivalent network
for $\tau'\geq\tau+\theta$ if its activation function $f$ is linear.
In previous publications \cite{Appeltant2011} such a network representation
was  derived for the case $\tau'=\tau$, however, in that case also
for nonlinear activation functions.

An important observation from Eq.~(\ref{eq:linear_network}) is that
the spectral radius of the matrix $A$ must be smaller than one, otherwise
the system \eqref{eq:linear_network} will not be asymptotically stable.
We achieve this by choosing appropriate parameter $\alpha=0.9$ (see
figure~\ref{fig:alpha} in Appendix~\ref{subsec:parameters}).

Equation~\eqref{eq:linear_network} allows us to calculate directly
some figures of merit in the following. We first use it to explain
the drops in the memory capacity in figure~\ref{fig:system_error}
for resonant delays. One important aspect to note, is that the basic
shape of equation~\eqref{eq:linear_network} does not change with
$\tau'$. Rather, a changing of the clock cycle leads to a change
of the evolution matrices $A$ and $W^{\mathrm{in}}$ of the equivalent
network. The obtained system~\eqref{eq:linear_network} can be equivalently
considered as a specific echo state network.

\section{Direct calculation of memory capacity}

\label{sec:mc}

One can find an estimation for the memory capacity of a reservoir
computing system by solving the system numerically and let it perform
the memory task. But there are also analytic methods for some cases.
In this section we explain how to calculate analytically the memory
capacity of the linear echo state network~\eqref{eq:linear_network}
which corresponds to the case $\tau'\geq \tau+\theta$.

Memory capacity was originally defined by Jaeger in \cite{Jaeger2002}.
In the following, we use a slightly modified formulation. Let the
elements $u(k)$ of the input sequence be independently $\mathcal{N}(0,1)$-distributed.
Jaeger introduced the quantity $\mathrm{MC}_{d}$ which indicates
how well the output $\hat{y}(k)$ of an ESN may approximate the input
value $u(k-d)$ which was fed into the reservoir $d$ time steps earlier.
The memory capacity for a recall of $d$ time steps in the past is
defined by 
\begin{align}
\mathrm{MC}_{d}:=\max_{W^{\mathrm{out}}}\left(1-\frac{\mathrm{E}[(W^{\mathrm{out}}X(k+d)-u(k))^{2}]}{\var(u(k))}\right),\label{eq:MC}
\end{align}
where $\mathrm{E}$ denotes the expectation value and we require the
initial state $X(0)$ of the reservoir to be stationary distributed
in order to ensure that this definition is consistent, i.e. that the
distribution of $X(k)$ does not depend on $k$. Since the spectral
radius of $A$ is less than one, the stationary distribution exists.
In such a case, the memory capacity~\eqref{eq:MC} with stationary
distributed $X(0)$ can be equivalently written as 
\begin{align}
\begin{split}\mathrm{MC}_{d} & :=\max_{W^{\mathrm{out}}}\left(1-\frac{\mathrm{E}[(W^{\mathrm{out}}X(d)-u(0))^{2}]}{\var(u(0))}\right)\\
 & \hphantom{:}=1-\min_{W^{\mathrm{out}}}\mathrm{E}[(W^{\mathrm{out}}X(d)-u(0))^{2}].\label{eq:MC_d}
\end{split}
\end{align}
Note that $u(0)\sim\mathcal{N}(0,1)$ means that we can drop the term
$\var(u(0))$ in~\eqref{eq:MC_d}. 

The total memory capacity $\mathrm{MC}$ is defined as the sum of
all $d$-step memory capacities 
\begin{align}
\mathrm{MC}:=\sum_{d=1}^{\infty}\mathrm{MC}_{d}.\label{eq:MC_sum}
\end{align}

In the following we denote the optimal output weight vector for \eqref{eq:MC_d}
by $W_{d}^{\mathrm{out}}$. Let $\Sigma$ be the covariance matrix
of the stationary distribution of the reservoir. Jaeger \cite{Jaeger2002}
noted that, if $\Sigma$ is invertible, one can apply the Wiener-Hopf
equation \cite{Haykin1995} to find 
\begin{align}
W_{d}^{\mathrm{out}}=(A^{d-1}W^{\mathrm{in}})^{\mathrm{T}}\Sigma^{-1}.\label{eq:MC_W_out}
\end{align}
For details we refer to Appendix \ref{subsec:mc}. Using this optimal
value $W_{d}^{\text{out}}$, the memory capacity~\eqref{eq:MC_d}
can be calculated as 
\begin{align}
\mathrm{MC}_{d}=(A^{d-1}W^{\mathrm{in}})^{\mathrm{T}}\Sigma^{-1}A^{d-1}W^{\mathrm{in}},\label{eq:MC_from_sigma}
\end{align}
where we have used the relations 
\begin{align}
\begin{split}\mathrm{E}[W_{d}^{\mathrm{out}}X(d)u(0)] & =\cov(W_{d}^{\mathrm{out}}X(d),u(0))\\
 & =(A^{d-1}W^{\mathrm{in}})^{\mathrm{T}}\Sigma^{-1}A^{d-1}W^{\mathrm{in}}
\end{split}
\end{align}
and 
\begin{align}
\begin{split}\mathrm{E}[(W_{d}^{\mathrm{out}}X(d))^{2}] & =\var(W_{d}^{\mathrm{out}}X(d))\\
 & =(A^{d-1}W^{\mathrm{in}})^{\mathrm{T}}\Sigma^{-1}A^{d-1}W^{\mathrm{in}}.
\end{split}
\end{align}
So once the covariance matrix of the reservoir $X$ is invertible,
one can directly calculate the memory capacity. The stationary distribution
of system~\eqref{eq:linear_network} with standard normal distributed
input elements $u(k)$ is a multivariate normal distribution with
mean zero and covariance matrix 
\begin{align}
\Sigma=\sum_{j=0}^{\infty}A^{j}W^{\mathrm{in}}(W^{\mathrm{in}})^{\mathrm{T}}A^{j\mathrm{T}}.\label{eq:Sigma}
\end{align}
We refer to Appendix~\ref{subsec:mc} for a derivation. It is worth to comment on the structure of the matrix $\Sigma$. We note that
the summands $A^{j}W^{\mathrm{in}}(W^{\mathrm{in}})^{\mathrm{T}}A^{j\mathrm{T}}$
in (\ref{eq:Sigma}) are rank one symmetric matrices with the norm
$\|A^{j}W^{\mathrm{in}}\|^{2}$. Since the spectral radius of $A$
is less than one, this norm converges to zero as $j\to\infty$ and there is only a finite
number of terms in (\ref{eq:Sigma}) which can make numerically significant
contribution to the rank of $\Sigma$. In addition, as we will see
in the following, these rank-one matrices may have almost coinciding eigenspaces. As a result, the matrix $\Sigma$ is in general numerically
not invertible. Since our approach for the derivation of equation~\eqref{eq:MC_from_sigma} relies on the invertibility of $\Sigma$, we cannot simply replace $\Sigma^{-1}$ by a pseudo-inverse. In order to obtain an invertible covariance matrix,
we need to perturb the stochastic process~\eqref{eq:linear_network}.
We choose a small number $\sigma_{\eta}>0$ and let $\eta(k)\sim\mathcal{N}(0,\Id)$
be a sequence of independent multivariate normal distributed random
variables. The stochastic process 
\begin{align}
X(k+1)=AX(k)+W^{\mathrm{in}}u(k)+\sigma_{\eta}\eta(k),\label{eq:linear_network_eta}
\end{align}
has the stationary distribution $\mathcal{N}(0,\Sigma_{\eta})$ where
the covariance matrix given by 
\begin{align}
\Sigma_{\eta}=\sum_{j=0}^{\infty}A^{j}(W^{\mathrm{in}}(W^{\mathrm{in}})^{\mathrm{T}}+\sigma_{\eta}\Id)A^{j\mathrm{T}}\label{eq:sigma_eta}
\end{align}
is invertible.

\section{Explanation for memory capacity gaps}

\label{sec:analytics}

Using the expressions~\eqref{eq:MC_sum} and \eqref{eq:MC_from_sigma}
for the memory capacity obtained in section~\ref{sec:mc}, we provide
an explanation for the loss of the memory capacity when $\tau'/\tau$
is close to rational numbers with small denominator. The explanation
is based on the structure of the covariance matrix $\Sigma_{\eta}$
given by equation~\eqref{eq:sigma_eta} and the corresponding expression
for the memory capacity, which we repeat here for convenience 
\begin{align}
\begin{split}\mathrm{MC} & =\sum_{d=1}^{\infty}\mathrm{MC}_{d},\\
\mathrm{MC}_{d} & =(A^{d-1}W^{\mathrm{in}})^{\mathrm{T}}\Sigma_{\eta}^{-1}A^{d-1}W^{\mathrm{in}},
\end{split}
\end{align}
where 
\begin{align}
\begin{split}\Sigma_{\eta} & :=\sum_{j=0}^{\infty}(\Pi_{j}+\sigma_{\eta}A^{j}A^{j\mathrm{T}}),\\
\Pi_{j} & :=A^{j}W^{\mathrm{in}}(W^{\mathrm{in}})^{\mathrm{T}}A^{j\mathrm{T}}.
\end{split}
\label{eq:sigmaeta}
\end{align}

Our further strategy is as follows: 
\begin{itemize}
\item[(i)] Firstly, we remark that the norms of the individual terms in the
sum~\eqref{eq:sigmaeta} are converging to zero due to the convergence
of the series. Hence, only the first finitely many terms play an important
role. For instance, for our previously chosen parameters in figure~\ref{fig:system_error},
the terms with $j\gtrsim30$ do not make a large contribution and
can be neglected. In the following we denote the approximate number
of significant terms by $j_{n}$. 
\item[(ii)] We show that the largest eigenvalue of the $j$-th term in \eqref{eq:sigmaeta}
can be approximated by $\|A^{j}W^{\text{in}}\|^{2}$ with the corresponding
eigenvector $A^{j}W^{\text{in}}$. 
\item[(iii)] We show that the memory capacity is high, i.e. $\mathrm{MC}_{d}\approx1$
for $d\leq j_{n}$, when the eigenvectors $A^{j}W^{\text{in}}$ corresponding
to the first relevant terms in the sum~\eqref{eq:sigmaeta} are orthogonal. 
\item[(iv)] Using our setup, we show numerically that the lower order resonances
$\tau'/\tau\approx a/b$, where $a,b\in\mathbb{N}$ and $b$ is small,
lead to the alignment of the eigenvectors $A^{j}W^{\text{in}}$, and
hence, to the loss of the memory capacity. The small shift from the exact
resonance values is explained by the standard drift property of delay
systems. 
\item[(v)] Finally we give an intuitive explanation of the obtained orthogonality
conditions. 
\end{itemize}

\paragraph{\textup{(i)} Convergence of the series~\eqref{eq:sigmaeta}}

The series \eqref{eq:sigmaeta} can be considered as the Neumann series
$(\mathrm{Id}-T)^{-1}=\sum_{j=0}^{\infty}T^{j},$ where $T\mathcal{X}:=A\mathcal{X}A^{\mathrm{T}}$,
applied to the matrix $W^{\mathrm{in}}(W^{\mathrm{in}})^{\mathrm{T}}+\sigma_{\eta}\mathrm{Id}$.
A sufficient condition for the convergence of such a series is that
$\|T^{k}\|_\mathrm{op}:=\sup_{\|\mathcal{X}\|=1}\|T^{k}\mathcal{X}\|<1$
for some $k>0$, where $\|\cdot\|$ is some matrix norm. Moreover,
$\|T^{k}\mathcal{X}\|=\|A^{k}\mathcal{X}A^{k\mathrm{T}}\|\leq\|A^{k}\|\|\mathcal{X}\|\|A^{k\mathrm{T}}\|=\|A^{k}\|^{2}$
when $\|\mathcal{X}\|=1$. Since the spectral radius of $A$ is smaller
than one, Gelfand's formula implies that there is a number $k>0$ such
that $\|A^{k}\|<1$, and hence, the sufficient condition $\|T^{k}\|_\mathrm{op}<1$
for the convergence of the series is satisfied.

\paragraph{\textup{(ii)} Estimating the largest eigenvalues and eigenvectors of the $j$-th
term in \eqref{eq:sigmaeta}}

Consider at first the term with $j=0$: $W^{\mathrm{in}}(W^{\mathrm{in}})^{\mathrm{T}}+\sigma_{\eta}\Id.$
The largest eigenvalue of this matrix is $\|W^{\mathrm{in}}\|_{2}^{2}+\sigma_{\eta}$
and the corresponding eigenvector is $W^{\mathrm{in}}$ as can be
easily checked by the direct calculation 
\begin{align}
[W^{\mathrm{in}}(W^{\mathrm{in}})^{\mathrm{T}}+\sigma_{\eta}\Id]W^{\mathrm{in}}=(\|W^{\mathrm{in}}\|_{2}^{2}+\sigma_{\eta})W^{\mathrm{in}}.
\end{align}
For all other eigenvectors $v$, which are orthogonal to $W^{\mathrm{in}}$
due to the symmetry of the matrix, the corresponding eigenvalues are
$\sigma_{\eta}$ because 
\begin{align}
[W^{\mathrm{in}}(W^{\mathrm{in}})^{\mathrm{T}}+\sigma_{\eta}\Id]v=W^{\mathrm{in}}\langle W^{\mathrm{in}},v\rangle+\sigma_{\eta}v=\sigma_{\eta}v.
\end{align}
These eigenvalues are by definition small, since $\sigma_{\eta}$
is a small perturbation.

We can also find approximations of the eigenvectors and eigenvalues
for the higher order terms $\Pi_{j}+\sigma_{\eta}A^{j}A^{j\mathrm{T}}$,
$j>0$. Namely, for the unperturbed matrix $\Pi_{j}$, the largest
eigenvalue is $\|A^{j}W^{\mathrm{in}}\|^{2}$ and the corresponding
eigenvector is $A^{j}W^{\mathrm{in}}$ because 
\begin{align}
\begin{split}\Pi_{j}A^{j}W^{\mathrm{in}} & =[A^{j}W^{\mathrm{in}}(W^{\mathrm{in}})^{\mathrm{T}}A^{j\mathrm{T}}]A^{j}W^{\mathrm{in}}\\
 & =A^{j}W^{\mathrm{in}}\langle A^{j}W^{\mathrm{in}},A^{j}W^{\mathrm{in}}\rangle\\
 & =\|A^{j}W^{\mathrm{in}}\|_{2}^{2}A^{j}W^{\mathrm{in}}.
\end{split}
\end{align}
All other eigenvalues are zero. Since the largest eigenvalue of $\Pi_{j}$
is geometrically and algebraically simple, it is continuous under
the perturbation by $\sigma_{\eta}\Id$. Hence, the largest eigenvalue
and the eigenvector of $\Pi_{j}+\sigma_{\eta}A^{j}A^{j\mathrm{T}}$
are approximated by $\|A^{j}W^{\mathrm{in}}\|^{2}$ and $A^{j}W^{\mathrm{in}}$
with an error of order $\sigma_{\eta}$. All other eigenvalues are
correspondingly small of order $\sigma_{\eta}$.

\paragraph{\textup{(iii)} The orthogonality of $A^{j}W^{\mathrm{in}}$ leads to the high
memory capacity}

Let $j_{n}$ be the number of terms in \eqref{eq:sigmaeta} that are
significant (see (i)), and let us assume that the eigenvectors $A^{j}W^{\mathrm{in}}$,
$j=0,1,\dots,j_{n}$ are close to be orthogonal, i.e. 
\begin{align}
\left|\left\langle A^{j}W^{\mathrm{in}},A^{i}W^{\mathrm{in}}\right\rangle \right|\ll1,\quad i,j=0,1,\dots,j_{n},\,\,\,j\neq i.\label{eq:orthog_asumption}
\end{align}
As we will see in (iv), such an assumption is indeed reasonable in
our setup. More precisely, one could consider \eqref{eq:orthog_asumption}
as $\left|\left\langle A^{j}W^{\mathrm{in}},A^{i}W^{\mathrm{in}}\right\rangle \right|<\varepsilon$
introducing another small parameter $\varepsilon\ll1$.

In case, when the orthogonality \eqref{eq:orthog_asumption} holds,
the largest eigenvalues of $\Sigma_{\eta}$ and their corresponding
eigenvectors can be approximated by $\|A^{j}W^{\mathrm{in}}\|^{2}$
and $A^{j}W^{\mathrm{in}}$, $j=0,\dots,j_{n}$. Indeed 
\begin{align}
\begin{split}\Sigma_{\eta}A^{j}W^{\mathrm{in}} & =\sum_{k=0}^{\infty}(\Pi_{k}A^{j}W^{\mathrm{in}}+\sigma_{\eta}A^{k}A^{k\mathrm{T}}A^{j}W^{\mathrm{in}})\\
 & =\|A^{j}W^{\mathrm{in}}\|_{2}^{2}A^{j}W^{\mathrm{in}}+\mathcal{O}(\sigma_{\eta})+\mathcal{O}(\varepsilon).
\end{split}
\end{align}
In this case, the memory capacity can be calculated as follows: 
\begin{align}
\begin{split} & \text{MC}_{d}=(A^{d-1}W^{\mathrm{in}})^{\mathrm{T}}\Sigma_{\eta}^{-1}A^{d-1}W^{\mathrm{in}}\\
 & =(A^{d-1}W^{\mathrm{in}})^{\mathrm{T}}\frac{1}{\|A^{d-1}W^{\mathrm{in}}\|_{2}^{2}}A^{d-1}W^{\mathrm{in}}+\mathcal{O}(\sigma_{\eta})+\mathcal{O}(\varepsilon)\\
 & =1+\mathcal{O}(\sigma_{\eta})+\mathcal{O}(\varepsilon)
\end{split}
\end{align}
for $d\leq j_{n}$. Hence, the orthogonality of the vectors $A^{j}W^{\mathrm{in}}$
with $A^{i}W^{\mathrm{in}}$, $i\ne j$ guarantees a high memory capacity.
We will present an intuitive explanation for this shortly.

\begin{figure}
\centering \includegraphics[width=255pt]{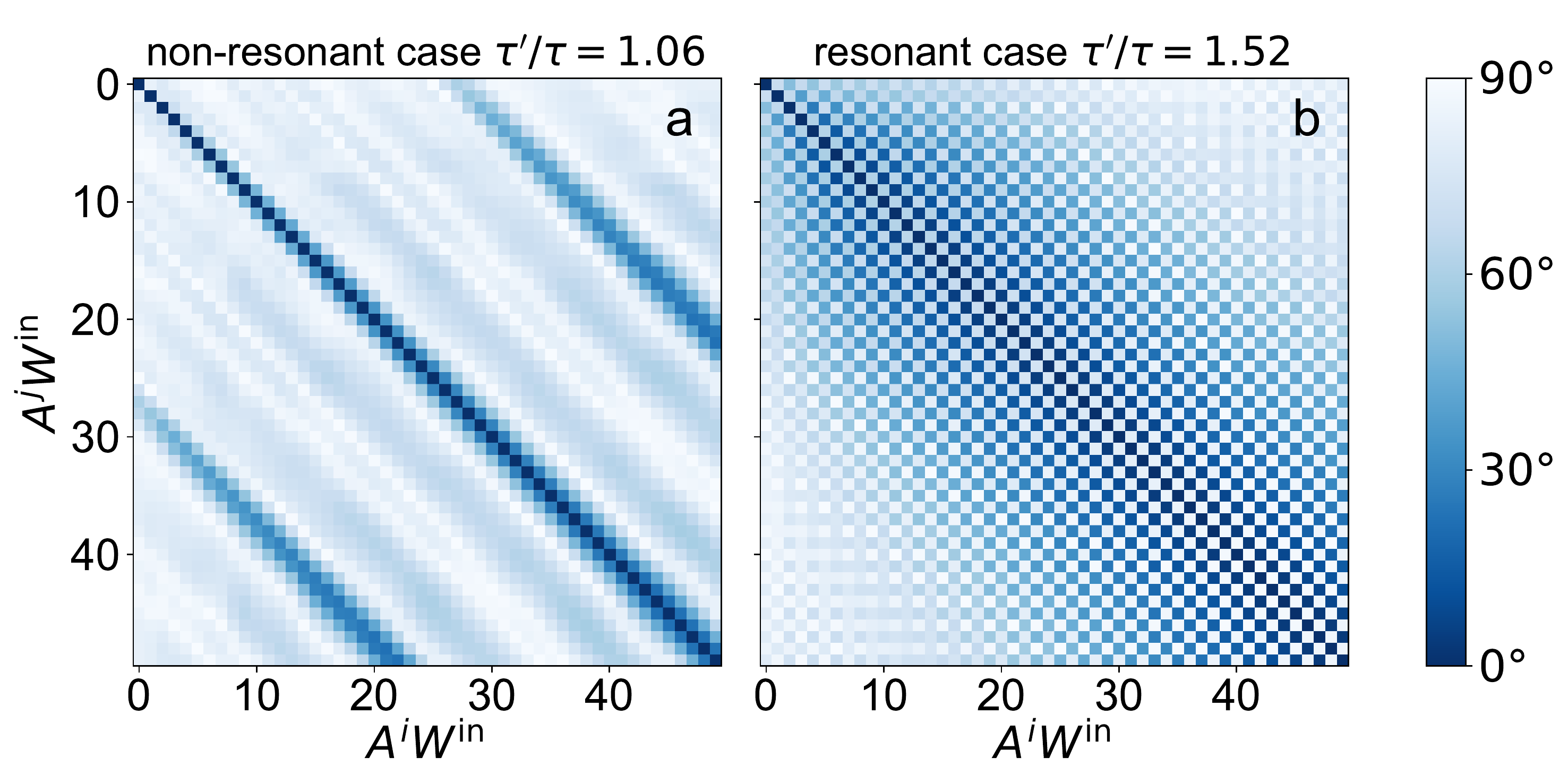} \caption{The angles between $A^{j}W^{\mathrm{in}}$ and $A^{i}W^{\mathrm{in}}$,
$i,j=1,\dots,50$ are plotted in color, measuring, in particular,
the orthogonality of the vectors $A^{j}W^{\mathrm{in}}$ with different
$j$. Panels (a) and (b) correspond to different ratios of $\tau'/\tau$:
(a) $\tau'/\tau=1.06$; (b) $\tau'/\tau=1.52\gtrsim3/2$. The former is off-resonant, while the later is close to the $3/2$ resonance, in particular when the drift property of the DDE system is taken into account. In (b) the
vectors $A^{j}W^{\mathrm{in}}$ point into the same direction for
$j$, $j+2$, $j+4$, etc., i.e. after two time steps the input values
$u(k)$ overlap in the state space of $X$ and the memory capacity
drops. In contrast, in the case $\tau'/\tau=1.06$ (a) it takes almost
30 time steps before the input overlaps with past inputs in the network
state. This explains the high memory capacity in this case, which
is illustrated in figure~\ref{fig:system_error}. }
\label{fig:angles} 
\end{figure}

\paragraph{\textup{(iv)} Resonances of $\tau'$ and $\tau$ lead to lower memory capacity}

The plots in figure~\ref{fig:angles} show $\left\langle A^{j}W^{\mathrm{in}},A^{i}W^{\mathrm{in}}\right\rangle $
for different ratios $\tau'/\tau$. White to light blue off-diagonal
squares indicate that assumption~\eqref{eq:orthog_asumption} is
satisfied, i.e. orthogonal or almost orthogonal vectors. Dark blue
indicates a strong parallelism of the vectors. As can bee seen in
the top panel of figure~\ref{fig:angles}, the assumption~\eqref{eq:orthog_asumption}
holds indeed for ratios $\tau'/\tau$ which yield a good memory performance.
Conversely, it is strongly violated for critical ratios $\tau'/\tau\gtrsim a/b$
with small denominator $b$, e.g. the center panel of figure~\ref{fig:angles}. 

We note that the critical ratios $\tau'/\tau$ are slightly shifted
from the exact resonant values $1$, $3/2$, etc. This shift is a manifestation
of the `drift' property of delay systems \cite{Yanchuk2017}, which
is caused by the fact that the effective round-trip of a signal in
a delay system equals to the delay $\tau$ plus a finite processing
time $\delta$ due to the integration (filtering). Therefore, the
small shift of the error peaks is actually due to a resonance between
the clock cycle $\tau'$ and $\tau+\delta$. This fact can be seen
also in the structure of the coupling matrix $A$ of the equivalent
network.

\paragraph{\textup{(v)} Intuitive explanations}

There is an additional intuitive understanding of the above derived
formulas. Recall that the original system of the reservoir of equation~\eqref{eq:delay_system}
combines the delay term $x(t-\tau)$ and the input $J(t)$ additively.
The approximated network formula for an equivalent network translated
this into the matrix $A$, which describes the free dynamics of the
network, and the driving term defined by $W^{\mathrm{in}}$. The state
of the network is given by an $N$-dimensional system, and thus can
at most hold $N$ orthogonal dimensions \cite{DAM12}. Each summand
of $\Sigma_{\eta}$ can now be understood as an imprint of the driving
term on the system after $j$ time steps. For $j=0$ the matrix $A^{0}=\Id$,
and thus the imprint is given by $W^{\mathrm{in}}$, i.e. the information
of the current step is stored in the nodes as given by the weights
of the effective input weight vector $W^{\mathrm{in}}$. In the next
step, the system will get an additional input, but also evolve according
to its local dynamics $A$. Thus, after one time step, the imprint
has transformed into $AW^{\mathrm{in}}$, i.e. the summand for $j=1$
and $\sigma_{\eta}\to0$. Now in every step, the information that
is currently present in the network will be `rotated' in the phase
space of the network according to $A$, while a new input will be
projected onto the direction of $W^{\mathrm{in}}$. This holds in
general, so that the $j$-th summand of $\Sigma_{\eta}$ of equation~\eqref{eq:sigma_eta}
$A^{j}W^{\mathrm{in}}$ describes the linear imprint of the input
$j$ steps in the past.

The orthogonality condition of equation~\eqref{eq:orthog_asumption}
then is the same concept as demanding that new information from the
inputs should not overwrite the already present information. If $A^{r}\approx s\Id$
for some $s\in\mathbb{R}$, then the information that was stored from
$r$ steps in the past will be partially overwritten by the currently
injected step and lost. Hence, ensuring that the orthogonality between
$A^{j}W^{\mathrm{in}}$ is fulfilled as much as possible will maximize
the linear memory. For the case of resonant feedback, i.e. $\tau'=\tau$,
this condition is not fulfilled. This is due to the fact, that $A$
has a strong diagonal component for the resonant cases, i.e. virtual
nodes are most strongly coupled to themselves. This is a simple consequence
of the fact that for $\tau=\tau'$, virtual nodes return to the single
real node at the same time that they are updated. Similarly, for higher
resonant cases $b\tau'=a\tau$, $A^{b}$ will in general have a strong
diagonal part and thus the eigenvector $A^{b}W^{in}$ will not be
orthogonal to $A^{0}W^{in}$, and the information will be overwritten.

\section{Discussion}

\label{sec:summary}

In this paper we have shown a generalization of the frequently used
time-delay reservoir computing for cases other than $\tau'=\tau$.
We observed that a sudden increase in the computing error (NARMA-10, Lorenz, and Santa-Fe
NRMSE) and a drop in the linear memory capacity ($\mathrm{MC}$) can
be seen for resonant cases of $b\tau'\approx a\tau$ with $a,b\in\mathbb{N}$
where $b$ is small for different activation functions including the
linear, $\tanh$, $\sin^{2}$, and the Mackey-Glass function. We derived
an equivalent network for the case $\tau' \geq \tau + \theta$ which extends the
previously studied case $\tau' = \tau$. Assuming a linear activation function $f(x)=\alpha x$,
we can analytically solve the resulting implicit equations and obtain
an expression for the total memory capacity $\mathrm{MC}$. Here we
find that the resulting memory capacity will be small for cases where
$\tau$ and $\tau'$ are resonant because the information within the
equivalent network will be overwritten by new inputs very quickly.
Even though our analytics so far are only derived for the linear case,
we expect these results to hold in more general situations, as numerical
simulations with several different nonlinear activation functions
in figures~\ref{fig:system_error}, \ref{fig:alpha}, \ref{fig:theta} indicate.
More detailed analysis can be performed in future studies.

\section*{Acknowledgements}

S.Y. acknowledges the financial support by the Deutsche Forschungsgemeinschaft
(DFG, German Research Foundation) - Project 411803875. A.R. and K.L.
acknowledge support from the Deutsche Forschungsgemeinschaft in the
framework of the CRC910. F.S. acknowledges financial support provided
by the Deutsche Forschungsgemeinschaft through the IRTG 1740. A.R. acknowledges the Spanish State Research Agency, through the Severo Ochoa and Maria de Maeztu Program for Centers and Units of Excellence in R\&D (MDM-2017-0711).

\section*{Appendix}

\global\long\def\thesubsection{\Alph{subsection}}%

\subsection{Derivation of equivalent networks}

\label{subsec:network}

This section presents a detailed derivation of the ESN represention
of TDRC systems. The derivation is structured as follows: 
\begin{enumerate}
\item The delay system~\eqref{eq:delay_system} is discretized such that
the state of a virtual node $x(k\tau'+n\theta)$ depends on the state
of its neighbor node $x(k\tau'+(n-1)\theta)$, the input $J_{k,n}$
and the state of a second node $x(k'\tau'+n'\theta)$ at the time
$k'\leq k$. In order to do so, we approximate the integral of the
continuous TDRC system on a small integration interval of length $\theta$
which covers the point $x(k'\tau'+n'\theta)$. 
\item The formulas for $n'$ and $k'$ are derived. 
\item The TDCR system can be written as a matrix equation. For this we use
the same vectorization \eqref{eq:vector_X} of $x(t)$ as for the
readout. An recurrent argument is employed to obtain the matrix equation. 
\item It follows that the discretized TDRC system can be represented by
an ESN if $\tau\leq\tau'-\theta$ and if the activation function $f$
is linear. 
\item For the sake of completeness, we formulate the equivalent ESN for
the classical case $\tau'=\tau$, which was described in \cite{Appeltant2011}. 
\end{enumerate}

\subsubsection{The delay reservoir system and discretization}

Consider the delay-system~\eqref{eq:delay_system}, which we repeat
here for convenience: 
\begin{align}
\dot{x}(t)=-x(t)+f[x(t-\tau)+\gamma J(t)],
\end{align}
where $\tau>0,\ \gamma>0$ and $f\colon\mathbb{R}\to\mathbb{R}$.

It follows that 
\begin{align}
e^{t-t_{0}}x(t)=x(t_{0})+\int_{t_{0}}^{t}e^{s-t_{0}}f[x(s-\tau)+\gamma J(s)]\intd s
\end{align}
for $t\geq t_{0}$. Set $t_{0}=k\tau'+(n-1)\theta$ and $t=k\tau'+n\theta$.
Then 
\begin{align}
\begin{split} & x(k\tau'+n\theta)=e^{-\theta}x(k\tau'+(n-1)\theta)\\
 & +\int_{0}^{\theta}e^{s-\theta}f[x(k\tau'+(n-1)\theta+s-\tau)+\gamma J_{k,n}]\intd s,\label{eq:integral}
\end{split}
\end{align}
where $J_{k,n}$ is defined in \eqref{eq:J_kn}. One option to discretize
the system, is to approximate the function $x$ by a step function
with step length $\theta$ which is constant on the integration interval.
One can find an appropriate step function by choosing $k'(k,n)$ and
$n'(n)$ such that 
\begin{align}
k'\tau'+n'\theta\in(k\tau'+(n-1)\theta-\tau,k\tau'+n\theta-\tau]\label{eq:kn_prime_in_interval}
\end{align}
and defining $x(t)\approx\tilde{x}(t):=x(k'\tau'+n'\theta)$ for $t\in(k\tau'+(n-1)\theta-\tau,k\tau'+n\theta-\tau]$.
Then, one can replace $x$ by $\tilde{x}$ in the integrand in equation~\eqref{eq:integral}.
This yields 
\begin{align}
\begin{split} & x(k\tau'+n\theta)\\
 & \approx e^{-\theta}x(k\tau'+(n-1)\theta)\\
 & \hphantom{{}=}+\int_{0}^{\theta}e^{s-\theta}f[x(k'(k,n)\tau'+n'(n)\theta)+\gamma J_{k,n}]\intd s\\
 & =e^{-\theta}x(k\tau'+(n-1)\theta)\\
 & \hphantom{{}=}+(1-e^{-\theta})f[x(k'(k,n)\tau'+n'(n)\theta)+\gamma J_{k,n}].
\end{split}
\label{eq:approx_integral}
\end{align}

\subsubsection{The choice of $k'$ and $n'$}

The floor and the ceiling function are denoted by $\lfloor\cdot\rfloor$
and $\lceil\cdot\rceil$, respectively. One can choose $k'$ and $n'$
in the following way:

\begin{figure}
\centering \includegraphics[width=255pt]{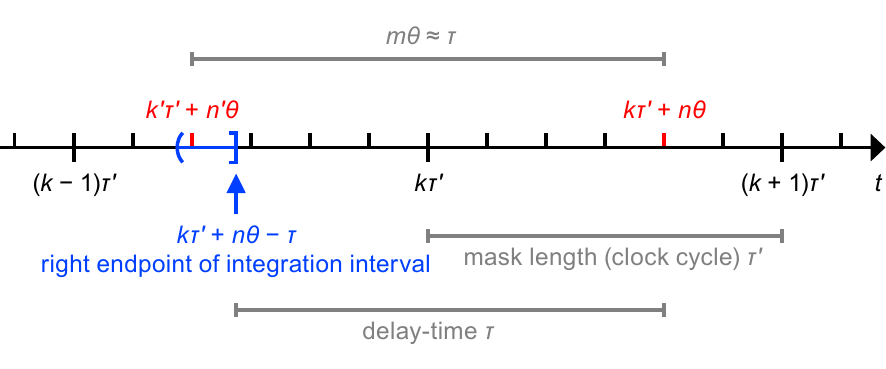} \caption{The time interval over which the function $x$ is integrated in equation~\eqref{eq:integral}
is highlighted in blue. As stated is equation~\eqref{eq:kn_prime_in_interval},
the point $k'\tau'+n'\theta$ must be chosen such that it lies within
this interval. In equation~\eqref{eq:approx_integral} the value
of $x$ on the integration interval is approximated by the value of
$x(k'\tau'+n'\theta)$. If the endpoints of the interval are grid
points, $k'\tau'+n'\theta$ is chosen to be the right endpoint.}
\label{fig:n_prime_k_prime} 
\end{figure}

First, let $m\in\mathbb{Z},\ m\geq1$ be the unique number such that
$\tau\in((m-1)\theta,m\theta]$, i.e. $m=\lceil\tau/\theta\rceil$.
Then 
\begin{align}
k'\tau'+n'\theta=k\tau'+n\theta-m\theta\label{eq:m_as_in_figure}
\end{align}
as illustrated in figure~\ref{fig:n_prime_k_prime}. Now, the choice
of $n'$ follows directly from the restriction $n'\in\{1,\ldots,N\}$.
It holds that 
\begin{align}
n'(n)=\begin{cases}
(n-m)\mod N, & \text{if }N\nmid(n-m),\\
N, & \text{if }N\mid(n-m).
\end{cases}\label{eq:n_prime}
\end{align}
From this result follows that 
\begin{align}
\begin{split}(n-m)\theta & =n'\theta+N\theta\cdot\begin{cases}
\left\lfloor \frac{n-m}{N}\right\rfloor , & \text{if }N\nmid(n-m),\\
\frac{n-m}{N}-1, & \text{if }N\mid(n-m),
\end{cases}\\
 & =n'\theta+\left(\left\lceil \frac{n-m}{N}\right\rceil -1\right)\tau'.
\end{split}
\end{align}
Hence, equation~\eqref{eq:m_as_in_figure} implies 
\begin{align}
k'(k,n)=k+\left\lceil \frac{n-m}{N}\right\rceil -1.\label{eq:k_prime}
\end{align}
Note that one has $k'=k$ as long as $n-m\in\{1,n-1\}$. If $n-m\in\{-N+1,\ldots,0\}$,
then $k'=k-1$. For $n-m\in\{-2N+1,\ldots,-N\}$ holds $k'=k-2$,
etc.

\subsubsection{Vectorization of the state space and a matrix equation for the discretized
system}

Define 
\begin{align}
X(k):=\begin{pmatrix}X_{1}(k)\\
\vdots\\
X_{N}(k)
\end{pmatrix}:=\begin{pmatrix}x((k-1)\tau'+\theta)\\
x((k-1)\tau'+2\theta)\\
\vdots\\
x((k-1)\tau'+N\theta)
\end{pmatrix}
\end{align}
and $\tilde{f}\equiv(1-e^{-\theta})f$. From \eqref{eq:approx_integral}
follows 
\begin{align}
\begin{split} & X_{1}(k+1)=x(k\tau'+\theta)\\
 & =e^{-\theta}x(k\tau')+\tilde{f}[x(k'(k,1)\tau'+n'(1)\theta)+\gamma J_{k,1}]\\
 & =e^{-\theta}X_{N}(k)+\tilde{f}[X_{n'(1)}(k'(k,1)+1)+\gamma J_{k,1}]
\end{split}
\end{align}
and repeated application of Eq.~\eqref{eq:approx_integral} yields
\begin{align}
\begin{split} & X_{n}(k+1)\\
 & =e^{-n\theta}X_{N}(k)\\
 & \hphantom{{}=}+e^{-(n-1)\theta}\tilde{f}[X_{n'(1)}(k'(k,1)+1)+\gamma J_{k,1}]\\
 & \hphantom{{}=}+e^{-(n-2)\theta}\tilde{f}[X_{n'(2)}(k'(k,2)+1)+\gamma J_{k,2}]\\
 & \hphantom{{}=}\ \ \vdots\\
 & \hphantom{{}=}+e^{-\theta}\tilde{f}[X_{n'(n-1)}(k'(k,n-1)+1)+\gamma J_{k,n-1}]\\
 & \hphantom{{}=}+\tilde{f}[X_{n'(n)}(k'(k,n)+1)+\gamma J_{k,n}]
\end{split}
\end{align}
for $n\in\{2,\ldots,N\}$.

These equations can by rewritten as a matrix equation. Let 
\begin{align}
A_{0}:=\begin{pmatrix}1 & 0 & \dots & 0\\
e^{-\theta} & 1 & \ddots & \vdots\\
\vdots & \ddots & \ddots & 0\\
e^{-(N-1)\theta} & \dots & e^{-\theta} & 1
\end{pmatrix}
\end{align}
and 
\begin{align}
\tilde{F}\begin{pmatrix}x_{1}\\
\vdots\\
x_{N}
\end{pmatrix}:=\begin{pmatrix}\tilde{f}(x_{1})\\
\vdots\\
\tilde{f}(x_{N})
\end{pmatrix}.
\end{align}
Then 
\begin{align}
\begin{split}X(k+1) & =A_{0}\tilde{F}\begin{pmatrix}X_{n'(1)}(k'(k,1)+1)+\gamma J_{k,1}\\
\vdots\\
X_{n'(N)}(k'(k,N)+1)+\gamma J_{k,N}
\end{pmatrix}\\
 & \hphantom{={}}+\begin{pmatrix}e^{-\theta}X_{N}(k)\\
\vdots\\
e^{-N\theta}X_{N}(k)
\end{pmatrix}\label{eq:matrix_system}
\end{split}
\end{align}

Let $\ell:=\lfloor m/N\rfloor$ and $q:=m\mod N$, as defined in \eqref{eq:mql_def},
i.e $m=\ell N+q$. By plugging this into equation~\eqref{eq:n_prime}
and noting that $1\leq n\leq N$ and $0\leq q\leq N-1$, one obtains
\begin{align}
n'(n)=\begin{cases}
n-q+N,\quad\text{for }n\leq q,\\
n-q,\quad\text{for }n>q,
\end{cases}
\end{align}
and by replacing $m$ by $\ell N+q$ equation~\eqref{eq:k_prime}
follows 
\begin{align}
k'(k,n)=\begin{cases}
k-\ell, & n>q,\\
k-\ell-1, & n\leq q.
\end{cases}
\end{align}
Hence, the vector $\left(X_{n'(n)}(k'(k,n)+1)\right)_{n=1,\ldots,N}$
can be written as follows: 
\begin{align}
\begin{split}
\begin{pmatrix}X_{n'(1)}(k'(k,1)+1)\\
\vdots\\
X_{n'(N)}(k'(k,N)+1)
\end{pmatrix}
&=\begin{pmatrix}0\\
\vdots\\
0\\
X_{1}(k+1-\ell)\\
\vdots\\
X_{N-q}(k+1-\ell)
\end{pmatrix} \\
&\hphantom{{}={}}+\begin{pmatrix}X_{N-q+1}(k-\ell)\\
\vdots\\
X_{N}(k-\ell)\\
0\\
\vdots\\
0
\end{pmatrix}.
\end{split}
\end{align}
Thus, the map~\eqref{eq:matrix_system} can be written as 
\begin{align}
\begin{split}\label{eq:matrix_system_shifted} & X(k+1):=A_{0}\tilde{F}[M_{q}X(k+1-\ell)\\
 & +M_{-(N-q)}X(k-\ell)+\gamma J_{k}]+A_{0}\begin{pmatrix}e^{-\theta}X_{N}(k)\\
0\\
\vdots\\
0
\end{pmatrix},
\end{split}
\end{align}
where the matrices $M_{q}=(\delta_{i,j+q})_{1\leq i,j\leq N}$ and
$M_{-(N-q)}=(\delta_{i,j-(N-q)})_{1\leq i,j\leq N}$ are shift matrices.

The matrix $A_{0}$ is invertible and can be used to transform the
system. Let $\tilde{X}:=A_{0}^{-1}X$. Then 
\begin{align}
\begin{split}\tilde{X}(k+1) & =B\tilde{X}(k)+\tilde{F}[A_{q}\tilde{X}(k+1-\ell)\\
 & \hphantom{{}=}+A_{-(N-q)}\tilde{X}(k-\ell)+\gamma J_{k}],\label{eq:network}
\end{split}
\end{align}
where the matrix 
\begin{multline}
A_{q}=M_{q}A_{0}=\\
\begin{pmatrix}0 & \cdots & \cdots & \cdots & \cdots & \cdots & 0\\
\vdots &  &  &  &  &  & \vdots\\
0 &  &  &  &  &  & \vdots\\
1 & 0 &  &  &  &  & \vdots\\
e^{-\theta} & 1 & \ddots &  &  &  & \vdots\\
\vdots & \ddots & \ddots & 0 &  &  & \vdots\\
e^{-(N-1-q)\theta} & \cdots & e^{-\theta} & 1 & 0 & \cdots & 0
\end{pmatrix}
\end{multline}
is obtained by a $q$ rows downwards shift of $A_{0}$ and the matrix
\begin{multline}
A_{-(N-q)}=M_{-(N-q)}A_{0}=\\
\begin{pmatrix}e^{-(N-q)\theta} & \cdots & e^{-2\theta} & e^{-\theta} & 1 & 0 & \cdots & 0\\
\vdots &  &  & \ddots & \ddots & \ddots & \ddots & \vdots\\
\vdots &  &  &  & \ddots & \ddots & \ddots & 0\\
e^{-(N-1)\theta} & \cdots & \cdots & \cdots & \cdots & e^{-2\theta} & e^{-\theta} & 1\\
0 & \cdots & \cdots & \cdots & \cdots & 0 & 0 & 0\\
\vdots &  &  &  &  &  &  & \vdots\\
0 & \cdots & \cdots & \cdots & \cdots & \cdots & \cdots & 0
\end{pmatrix}
\end{multline}
is obtained by an $N-q$ rows upwards shift of $A_{0}$ and 
\begin{align}
B=\begin{pmatrix}e^{-N\theta} & \dots & e^{-\theta}\\
0 & \dots & 0\\
\vdots &  & \vdots\\
0 & \dots & 0
\end{pmatrix}.\label{eq:B}
\end{align}
The equation \eqref{eq:B} for matrix $B$ follows from equation~\eqref{eq:matrix_system_shifted}.
It must hold that 
\begin{equation}
\begin{pmatrix}e^{-\theta}X_{N}(k)\\
0\\
\vdots\\
0
\end{pmatrix}=B\tilde{X}(k)=BA_{0}^{-1}X(k).
\end{equation}
Hence, 
\begin{equation}
B=\begin{pmatrix}0 & \dots & 0 & e^{-\theta}\\
0 & \dots & \dots & 0\\
\vdots &  &  & \vdots\\
0 & \dots & \dots & 0
\end{pmatrix}A_{0}.
\end{equation}

\subsubsection{An ESN representation of TDRC systems with suitable parameters}

\label{subsec:linear_network}

\begin{figure}
\centering \includegraphics[width=255pt]{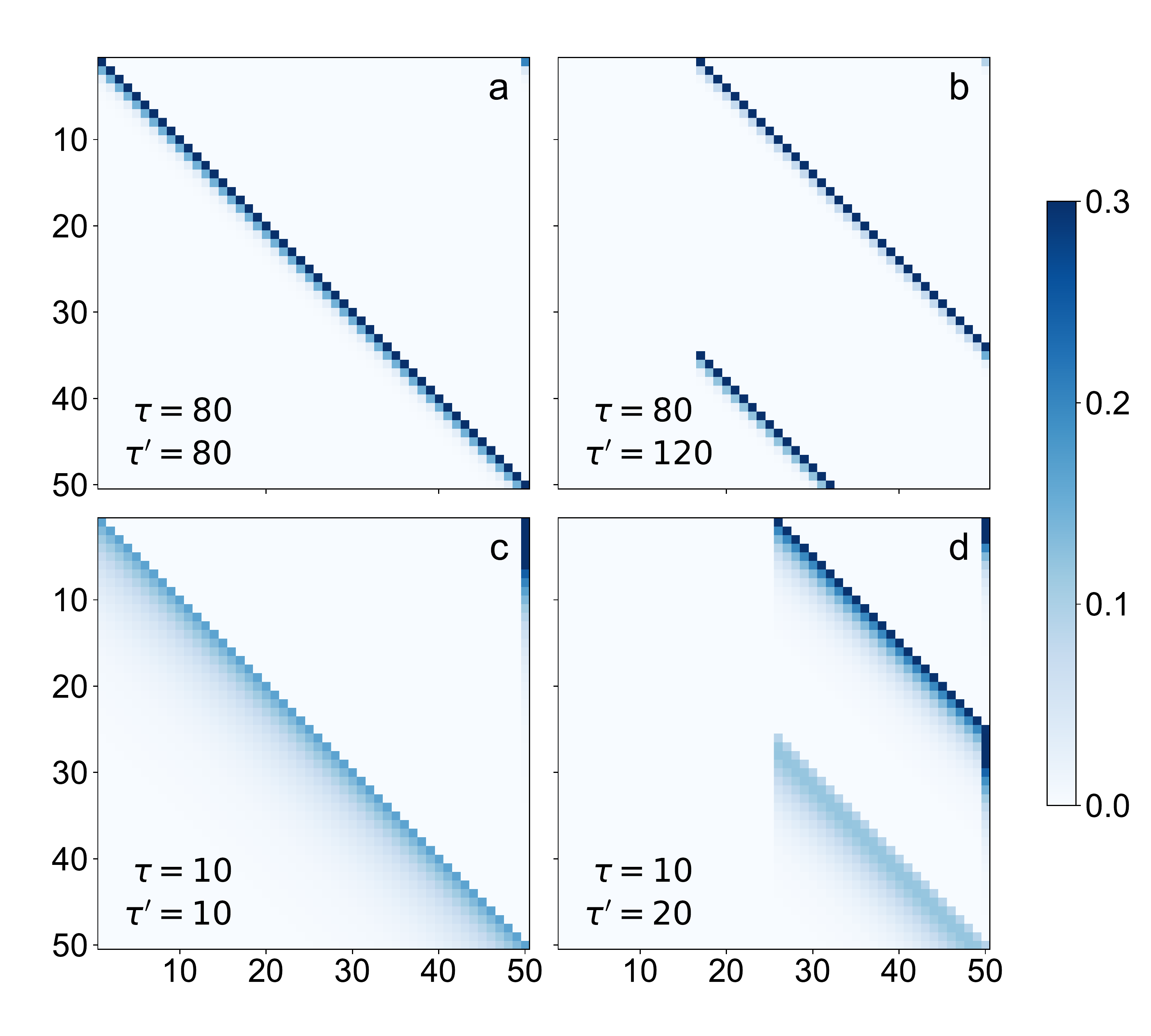}
\caption{Plot of the network matrix $A$, given by \eqref{eq:network_matrix_A}
resp. \eqref{eq:network_matrix_A_classical}, for $N=50$ and diverse values of $\tau$ and $\tau'$. The connection weights are truncated at $0.3$. The panels (a) and (c) show matrices for the classical case $\tau = \tau'$. (See subsection~\ref{subsec:classical_network}.)
The panels (b) and (d) show matrices for the case $\tau\protect\leq\tau'-\theta$
(subsection~\ref{subsec:linear_network}). These particular examples have resonant values of $\tau$ and $\tau'$, in fact we have (b) $\tau'/\tau = 1.5$ and (d) $\tau'/\tau = 2$.
The strongest connection
weights lie on diagonal lines which are shifted as the ratio of $\tau'$
and $\tau$ changes. The weights below these lines scale with the
factor $e^{-n\theta}$, where $n$ is the distance to the line. Since
$\theta=\tau'/N$, the (off-diagonal) weights are larger in panels
(c) and (d), were $\tau'$ is smaller.}
\label{fig:network} 
\end{figure}

If $\tau\leq\tau'-\theta$, then $\ell=0$. This follows from the
definitions $\ell:=\lfloor m/N\rfloor$ and $m=\lceil\tau/\theta\rceil$.
Equation~\eqref{eq:network} is in this case an implicit map: 
\begin{align}
\begin{split}\tilde{X}(k+1) & =B\tilde{X}(k)\\
 & \hphantom{={}}+\tilde{F}[A_{q}\tilde{X}(k+1)+A_{-(N-q)}\tilde{X}(k)+\gamma J_{k}].
\end{split}
\end{align}
However, for a linear activation function $f(x)=\alpha x$, where
$\alpha$ is a scalar, holds $\tilde{f}(x)=(1-e^{-\theta})\alpha x$
and hence one obtains the explicit linear map 
\begin{align}
\begin{split}\tilde{X}(k+1) & =(\Id-\nu\alpha A_{q})^{-1}(B+\nu\alpha A_{-(N-q)})\tilde{X}(k)\\
 & \hphantom{{}=}+\nu\alpha\gamma(\Id-\nu\alpha A_{q})^{-1}J_{k},
\end{split}
\end{align}
where $\nu:=1-e^{-\theta}$. Since $\tilde{X}=A_{0}^{-1}X$ and $J_{k}=W^{\mathrm{mask}}u_{k}$,
one can write this map in the original coordinates and in terms of
the original input sequence 
\begin{align}
X(k+1)=AX(k)+Wu(k),\label{eq:linear}
\end{align}
where 
\begin{align}
A:=A_{0}(\Id-\nu\alpha A_{q})^{-1}(B+\nu\alpha A_{-(N-q)})A_{0}^{-1}\label{eq:network_matrix_A}
\end{align}
and 
\begin{align}
W^{\mathrm{in}}:=\nu\alpha\gamma A_{0}(\Id-\nu\alpha A_{q})^{-1}W^{\mathrm{mask}}.
\end{align}
The network matrix A is plotted in figure~\ref{fig:network} for
different parameters.

\subsubsection{The ESN representation of classical TDRC systems}

\label{subsec:classical_network}

The article \cite{Appeltant2011} contains a description of an equivalent
echo state network for TDRC systems with $\tau'=\tau$. This description
is consistent with the case $\tau\in(\tau'-\theta,\tau']$ in the
framework of our discretization. In this case, 
\begin{align}
m=N,\quad\ell=1,\quad q=0,
\end{align}
and therefore, $A_{q}=A_{0}$ and $A_{-(N-q)}$ is the zero matrix.
Thus, equation~\eqref{eq:network} simplifies to 
\begin{align}
\tilde{X}(k+1)=B\tilde{X}(k)+\tilde{F}[A_{0}\tilde{X}(k)+\gamma J_{k}].\label{eq:network_classical}
\end{align}
For a linear activation function $f(x)=\alpha x$ and $\tau\leq\tau'-\theta$,
i.e. $\ell=0$, the equivalent network written in the original coordinates
is 
\begin{align}
X(k+1)=AX(k)+Wu(k),\label{eq:linear_classical}
\end{align}
where 
\begin{align}
A:=A_{0}BA_{0}^{-1}+\nu\alpha A_{0}\label{eq:network_matrix_A_classical}
\end{align}
and 
\begin{align}
W^{\mathrm{in}}:=\nu\alpha\gamma A_{0}W^{\mathrm{mask}}.
\end{align}

\subsection{Derivation of the memory capacity formula}

\label{subsec:mc}

We consider the linear echo state network 
\begin{align}
X(k+1)=AX(k)+W^{\mathrm{in}}u(k),
\end{align}
where the input elements $u(k)$ are independently $\mathcal{N}(0,1)$-distributed.
In section~\ref{sec:mc} we defined 
\begin{align}
\mathrm{MC}_{d}=\max_{W^{\mathrm{out}}}\left(1-\E[(W^{\mathrm{out}}X(k+d)-u(k))^{2}]\right)\label{eq:mc_d2}
\end{align}
and we claimed that 
\begin{align}
W_{d}^{\mathrm{out}}=(A^{d-1}W^{\mathrm{in}})^{\mathrm{T}}\Sigma^{-1}
\end{align}
is the optimal argument for \eqref{eq:mc_d2}. In the following we
show that $W_{d}^{\mathrm{out}}$ is indeed the optimal argument for
\eqref{eq:mc_d2}.

In order to maximize \eqref{eq:mc_d2}, we need to minimize the mean
square error 
\begin{align}
\begin{split}\MSE & =\E[(W^{\mathrm{out}}X(k+d)-u(k))^{2}]\\
 & =\E[(W^{\mathrm{out}}X(k+d))^{2}]+\E[u(k)^{2}]\\
 & \hphantom{{}=}-2\E[W^{\mathrm{out}}X(k+d)u(k)].
\end{split}
\end{align}
We know that $X(k)\sim\mathcal{N}(0,\Sigma)$ and hence 
\begin{align}
W^{\mathrm{out}}X(k)\sim\mathcal{N}(0,W^{\mathrm{out}}\Sigma(W^{\mathrm{out}})^{\mathrm{T}}).
\end{align}
Note that $W^{\mathrm{out}}\Sigma(W^{\mathrm{out}})^{\mathrm{T}}$
is a scalar because $W^{\mathrm{out}}$ is a row vector. Since the
mean of $W^{\mathrm{out}}X(k+d)$ is zero and $u(k)\sim\mathcal{N}(0,1)$,
we have 
\begin{align}
\begin{split}\E[(W^{\mathrm{out}}X(k+d))^{2}] & =\var(W^{\mathrm{out}}X(k+d))\\
 & =W^{\mathrm{out}}\Sigma(W^{\mathrm{out}})^{\mathrm{T}},
\end{split}
\\
\E[u(k)^{2}] & =1,\\
\E[W^{\mathrm{out}}X(k+d)u(k)] & =\cov(u(k),W^{\mathrm{out}}X(k+d)).
\end{align}
Moreover, 
\begin{align}
\begin{split} & W^{\mathrm{out}}X(k+d)\\
 & =W^{\mathrm{out}}\left(A^{d}X(k)+\sum_{j=0}^{d-1}A^{j}W^{\mathrm{in}}u(k+d-1-j)\right)
\end{split}
\end{align}
and $u(k)$ is independent of $X(k)$. Therefore, 
\begin{align}
\begin{split} & \cov(u(k),W^{\mathrm{out}}X(k+d))\\
 & =\cov(u(k),W^{\mathrm{out}}A^{d-1}W^{\mathrm{in}}u(k))\\
 & =W^{\mathrm{out}}A^{d-1}W^{\mathrm{in}}.
\end{split}
\end{align}
Thus, we obtain 
\begin{align}
\MSE=W^{\mathrm{out}}\Sigma(W^{\mathrm{out}})^{\mathrm{T}}+1-2W^{\mathrm{out}}A^{d-1}W^{\mathrm{in}}.
\end{align}
Since the mean square error is quadratic in the argument $W^{\mathrm{out}}=(w_{1}^{\mathrm{out}},\ldots,w_{N}^{\mathrm{out}})$,
it has exactly one local minimum, which is the global minimum. A row
vector $W_{d}^{\mathrm{out}}$ is the minimum argument if and only
if 
\begin{align}
\frac{\partial}{\partial w_{n}^{\mathrm{out}}}\MSE(W_{d}^{\mathrm{out}})=0,\quad n=1,\ldots,N.
\end{align}

For a quadratic form 
\begin{align}
Q(v)=v^{\mathrm{T}}Mv,
\end{align}
where $v\in\mathbb{R}^{N}$ and $M$ is a symmetric matrix, the vector
of the partial derivatives is given by 
\begin{align}
\frac{\partial Q(v)}{\partial v}=2v^{\mathrm{T}}M.
\end{align}
Therefore, 
\begin{align}
\frac{\partial\MSE}{\partial W^{\mathrm{out}}}=2W^{\mathrm{out}}\Sigma-2(A^{d-1}W^{\mathrm{in}})^{\mathrm{T}}
\end{align}
and hence 
\begin{align}
W_{d}^{\mathrm{out}}\Sigma=(A^{d-1}W^{\mathrm{in}})^{\mathrm{T}}.
\end{align}
This formula is called Wiener-Hopf equation \cite{Haykin1995}. It
follows that 
\begin{align}
W_{d}^{\mathrm{out}}=(A^{d-1}W^{\mathrm{in}})^{\mathrm{T}}\Sigma^{-1}.
\end{align}

\subsection{The NARMA-10 benchmark}

\label{subsec:narma}

The 10th-order nonlinear autoregressive moving average (NARMA-10)
task was introduced in \cite{Atiya2000} to evaluate the performance
of machine learning methods on time series estimation. The NARMA-10
sequence $(y(k))_{k\geq0}$ is defined as follows: for an input sequence
with independently $\mathcal{U}(0,0.5)$-distributed elements $u(k)$,
let 
\begin{align}
y(0)=y(1)=\ldots=y(9)=0
\end{align}
and 
\begin{align}
\begin{split}y(k+1) & =0.3y(k)+0.05y(k)\left(\sum_{j=0}^{9}y(k-j)\right)\\
 & \hphantom{{}=}+1.5u(k-9)u(k)+0.1
\end{split}
\end{align}
for $k\geq9$.

In order to evaluate the performance of a reservoir computer, we choose
sufficiently large numbers $k_{0},K\in\mathbb{N}$ and we compare
the output values $\hat{y}(k_{0}+1),\ldots,\hat{y}(k_{0}+K)$ to the
desired target values $y(k_{0}+1),\ldots,y(k_{0}+K)$ by the normalized
mean square error 
\begin{align}
\mathrm{NMSE}=\frac{1}{K}\sum_{k=k_{0}+1}^{k_{0}+K}\frac{(\hat{y}(k)-y(k))^{2}}{\var(y)}.\label{eq:NMSE}
\end{align}

\subsection{The Lorenz benchmark}

\label{subsec:lorenz}

For the Lorenz task we use the three-dimensional Lorenz system: 
\begin{align}
\begin{split}\dot{\xi} & =10(\upsilon-\zeta),\\
\dot{\upsilon} & =\xi(28-\zeta)-\upsilon,\\
\dot{\zeta} & =\xi\upsilon-\frac{8}{3}\zeta.
\end{split}
\end{align}
We obtain a three-dimensional input sequence $u(k)$ of the reservoir
by sampling with period $0.1$ and normalization of all components,
i.e. 
\begin{align}
u(k):=\begin{pmatrix}\xi(k/10)/\var(\xi)\\
\upsilon(k/10)/\var(\upsilon)\\
\zeta(k/10)/\var(\zeta)
\end{pmatrix}.
\end{align}
The task is a one-time-step-prediction task for the $\xi$-component,
i.e. the target sequence is given by $y(k):=\xi((k+1)/10)/\var(\xi)$.
For the evaluation we use the NMSE \eqref{eq:NMSE}.

\subsection{The Santa Fe benchmark}

\label{subsec:santafe}

For the Santa Fe time-series prediction task we use a normalized version
of the Santa Fe laser series \cite{Weigend1993} as input. The target
is to predict the next value of the series, i.e. as the Lorenz task,
the Santa Fe task is a one-time-step-prediction task. For the evaluation
we use the NMSE \eqref{eq:NMSE}.

\subsection{Parameters}
\label{subsec:parameters}

Our choice of parameters does not significantly influence our results. In this section we present the numerical simulations that we used to verify this for $\alpha$, $\theta$ resp. $N$. 

\begin{figure}
	\centering \includegraphics[width=255pt]{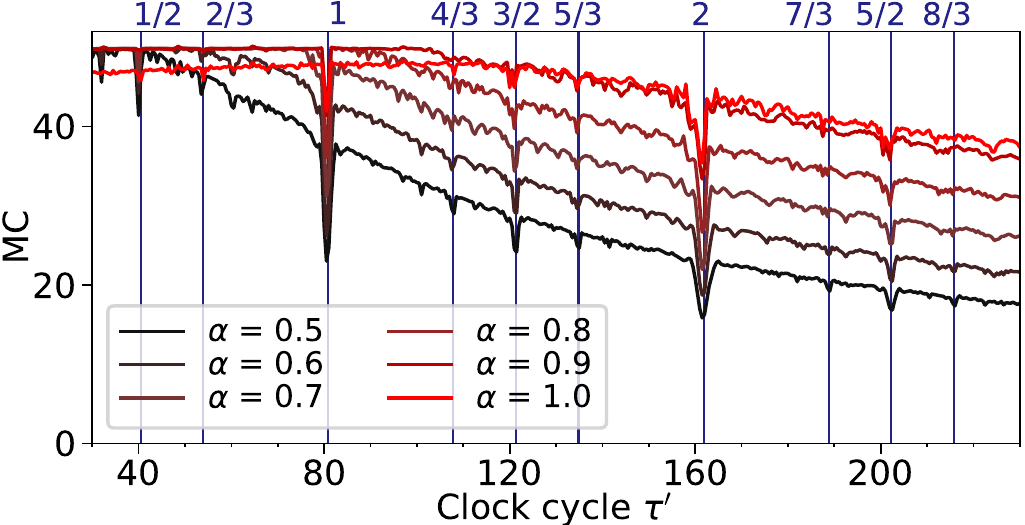}
	\caption{Influence of the parameter $\alpha$ of the linear model of Eq.~\eqref{eq:f_linear} on the linear memory capacity. All other parameters are as in table~\ref{table:parameters}. The numerical simulations in the main part of this report were obtained with $\alpha = 0.9$.} 
	\label{fig:alpha} 
\end{figure}

Figure~\ref{fig:alpha} shows the memory capacity as a function of the clock cycle $\tau'$ for different $\alpha$. The effect is visible for a large range of values. In the paper, we have used $\alpha=0.9$, corresponding to the highest capacity out of the tested ones. 

\begin{figure}
	\centering \includegraphics[width=255pt]{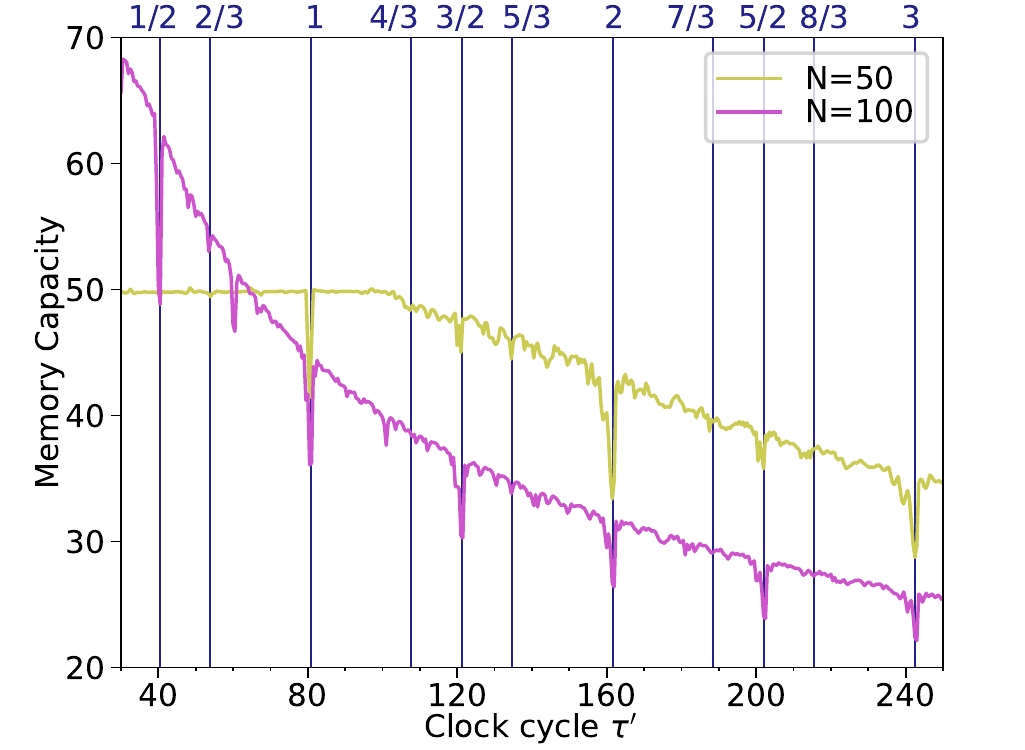}
	\caption{Influence of the parameter $N$ for the linear model of Eq.~\eqref{eq:f_linear} on the linear memory capacity. All other parameters are as in table~\ref{table:parameters}. The numerical simulations in the main part of this report were obtained with $N = 50$.}
	\label{fig:theta} 
\end{figure}

Figure~\ref{fig:theta} shows the memory capacity as a function of the clock cycle $\tau'$ for two different numbers of virtual nodes $N= 50$ and $N=100$. Since the time per virtual node is $\theta = \tau' / N$, it ranges between $0.6$ and $5.0$ in the case $N = 50$ and between $0.3$ and $2.5$ in the case $N = 100$. In both cases The effect is visible. In the paper, we have used $N=50$. 

\bibliographystyle{elsarticle-num}
\bibliography{preprint}

\begin{thebibliography}{10}
\expandafter\ifx\csname url\endcsname\relax
  \def\url#1{\texttt{#1}}\fi
\expandafter\ifx\csname urlprefix\endcsname\relax\def\urlprefix{URL }\fi
\expandafter\ifx\csname href\endcsname\relax
  \def\href#1#2{#2} \def\path#1{#1}\fi

\bibitem{Jaeger2001}
H.~Jaeger, {The ``echo state'' approach to analysing and training recurrent
  neural networks}, Ger. Natl. Res. Cent. Inf. Technol. GMD Tech. Rep. 148.

\bibitem{Maass2002}
W.~Maass, T.~Natschl{\"{a}}ger, H.~Markam, {Real-time computing without stable
  states: A new framework for neural computation based on perturbations},
  Neural Comput. 14~(11) (2002) 2531--2560.

\bibitem{SAN17a}
G.~Van~der Sande, D.~Brunner, M.~C. Soriano, Advances in photonic reservoir
  computing, Nanophotonics 6~(3) (2017) 561.
\newblock \href {http://dx.doi.org/https://doi.org/10.1515/nanoph-2016-0132}
  {\path{doi:https://doi.org/10.1515/nanoph-2016-0132}}.

\bibitem{BRU18a}
D.~Brunner, B.~Penkovsky, B.~A. Marquez, M.~Jacquot, I.~Fischer, L.~Larger,
  Tutorial: Photonic neural networks in delay systems, J. Appl. Phys. 124~(15)
  (2018) 152004.
\newblock \href {http://dx.doi.org/10.1063/1.5042342}
  {\path{doi:10.1063/1.5042342}}.

\bibitem{Jaeger2004}
H.~Jaeger, H.~Haas, {Harnessing Nonlinearity: Predicting Chaotic Systems and
  Saving Energy in Wireless Communication}, Science (80-. ). 304~(5667) (2004)
  78--80.
\newblock \href {http://dx.doi.org/10.1126/SCIENCE.1091277}
  {\path{doi:10.1126/SCIENCE.1091277}}.

\bibitem{Verstraeten2005}
D.~Verstraeten, B.~Schrauwen, D.~Stroobandt, J.~{Van Campenhout},
  \href{https://www.sciencedirect.com/science/article/pii/S0020019005001523?via{\%}3Dihub}{{Isolated
  word recognition with the Liquid State Machine: a case study}}, Inf. Process.
  Lett. 95~(6) (2005) 521--528.
\newblock \href {http://dx.doi.org/10.1016/J.IPL.2005.05.019}
  {\path{doi:10.1016/J.IPL.2005.05.019}}.
\newline\urlprefix\url{https://www.sciencedirect.com/science/article/pii/S0020019005001523?via{\%}3Dihub}

\bibitem{Verstraeten2006}
D.~Verstraeten, B.~Schrauwen, D.~Stroobandt,
  \href{http://ieeexplore.ieee.org/document/1716215/}{{Reservoir-based
  techniques for speech recognition}}, in: 2006 IEEE Int. Jt. Conf. Neural
  Netw. Proc., IEEE, 2006, pp. 1050--1053.
\newblock \href {http://dx.doi.org/10.1109/IJCNN.2006.246804}
  {\path{doi:10.1109/IJCNN.2006.246804}}.
\newline\urlprefix\url{http://ieeexplore.ieee.org/document/1716215/}

\bibitem{ARG18}
A.~Argyris, J.~Bueno, I.~Fischer, Photonic machine learning implementation for
  signal recovery in optical communications, Sci. Rep. 8~(8487) (2018) 1--13.
\newblock \href {http://dx.doi.org/10.1038/s41598-018-26927-y}
  {\path{doi:10.1038/s41598-018-26927-y}}.

\bibitem{PAT18}
J.~Pathak, B.~Hunt, M.~Girvan, Z.~Lu, E.~Ott, {Model-Free Prediction of Large
  Spatiotemporally Chaotic Systems from Data: A Reservoir Computing Approach},
  Phys. Rev. Lett. 120~(2) (2018) 24102.
\newblock \href {http://dx.doi.org/10.1103/physrevlett.120.024102}
  {\path{doi:10.1103/physrevlett.120.024102}}.

\bibitem{Jaeger2002}
H.~Jaeger, {Short Term Memory in Echo State Networks}, Ger. Natl. Res. Cent.
  Inf. Technol. GMD Tech. Rep. 152.

\bibitem{Lymburn2019}
T.~Lymburn, A.~Khor, T.~Stemler, D.~C. Corr\~{e}a, M.~Small, T.~J\"ungling,
  {Consistency in echo-state networks}, Chaos 29.

\bibitem{Appeltant2011}
L.~Appeltant, M.~Soriano, G.~{Van der Sande}, J.~Danckaert, S.~Massar,
  J.~Dambre, B.~Schrauwen, C.~Mirasso, I.~Fischer,
  \href{http://www.nature.com/articles/ncomms1476}{{Information processing
  using a single dynamical node as complex system}}, Nat. Commun. 2~(1) (2011)
  468.
\newblock \href {http://dx.doi.org/10.1038/ncomms1476}
  {\path{doi:10.1038/ncomms1476}}.
\newline\urlprefix\url{http://www.nature.com/articles/ncomms1476}

\bibitem{ROE18a}
A.~R{\"{o}}hm, K.~L{\"{u}}dge, {Multiplexed networks: reservoir computing with
  virtual and real nodes}, J. Phys. Commun. 2 (2018) 85007.
\newblock \href {http://dx.doi.org/10.1088/2399-6528/aad56d}
  {\path{doi:10.1088/2399-6528/aad56d}}.

\bibitem{Erneux2009}
T.~Erneux, {Applied Delay Differential Equations}, Vol.~3 of Surveys and
  Tutorials in the Applied Mathematical Sciences, Springer, 2009.

\bibitem{Erneux2017}
T.~Erneux, J.~Javaloyes, M.~Wolfrum, S.~Yanchuk,
  \href{http://aip.scitation.org/doi/10.1063/1.5011354}{{Introduction to Focus
  Issue: Time-delay dynamics}}, Chaos An Interdiscip. J. Nonlinear Sci. 27~(11)
  (2017) 114201.
\newblock \href {http://dx.doi.org/10.1063/1.5011354}
  {\path{doi:10.1063/1.5011354}}.
\newline\urlprefix\url{http://aip.scitation.org/doi/10.1063/1.5011354}

\bibitem{Yanchuk2017}
S.~Yanchuk, G.~Giacomelli,
  \href{http://stacks.iop.org/1751-8121/50/i=10/a=103001?key=crossref.f760c062e912b820ac69c9174ac61305}{{Spatio-temporal
  phenomena in complex systems with time delays}}, J. Phys. A Math. Theor.
  50~(10) (2017) 103001.
\newblock \href {http://dx.doi.org/10.1088/1751-8121/50/10/103001}
  {\path{doi:10.1088/1751-8121/50/10/103001}}.
\newline\urlprefix\url{http://stacks.iop.org/1751-8121/50/i=10/a=103001?key=crossref.f760c062e912b820ac69c9174ac61305}

\bibitem{Brunner2013}
D.~Brunner, M.~C. Soriano, C.~R. Mirasso, I.~Fischer, {Parallel photonic
  information processing at gigabyte per second data rates using transient
  states}, Nat. Commun. 2.

\bibitem{Larger2012}
L.~Larger, M.~C. Soriano, D.~Brunner, L.~Appeltant, J.~M. Gutierrez,
  L.~Pesquera, C.~R. Mirasso, I.~Fischer,
  \href{http://www.opticsexpress.org/abstract.cfm?URI=oe-20-3-3241}{{Photonic
  information processing beyond Turing: an optoelectronic implementation of
  reservoir computing}}, Opt. Express 20~(3) (2012) 3241--3249.
\newblock \href {http://dx.doi.org/10.1364/OE.20.003241}
  {\path{doi:10.1364/OE.20.003241}}.
\newline\urlprefix\url{http://www.opticsexpress.org/abstract.cfm?URI=oe-20-3-3241}

\bibitem{Kuriki2018}
Y.~Kuriki, J.~Nakayama, K.~Takano, A.~Uchida, {Impact of input mask signals on
  delay-based photonic reservoir computing with semiconductor lasers}, Optics
  Express 26.

\bibitem{Rulkov1995}
N.~F. Rulkov, M.~M. Sushchik, L.~S. Tsimring, H.~D.~I. Abarbanel,
  \href{https://link.aps.org/doi/10.1103/PhysRevE.51.980}{Generalized
  synchronization of chaos in directionally coupled chaotic systems}, Phys.
  Rev. E 51 (1995) 980--994.
\newblock \href {http://dx.doi.org/10.1103/PhysRevE.51.980}
  {\path{doi:10.1103/PhysRevE.51.980}}.
\newline\urlprefix\url{https://link.aps.org/doi/10.1103/PhysRevE.51.980}

\bibitem{Pikovsky2001}
A.~Pikovsky, M.~Rosenblum, J.~Kurths, {Synchronization: A universal concept in
  nonlinear science}, Cambridge University Press, 2001.

\bibitem{Hale1993}
J.~Hale, S.~Verduyn~Lunel, An Introduction to Functional Differential
  Equations, Vol.~99, 1993.

\bibitem{Smith2010}
H.~Smith, An Introduction to Delay Differential Equations with Applications to
  the Life Sciences, Vol.~57, 2010.
\newblock \href {http://dx.doi.org/10.1007/978-1-4419-7646-8}
  {\path{doi:10.1007/978-1-4419-7646-8}}.

\bibitem{Schumacher2013}
J.~Schumacher, H.~Toutounji, G.~Pipa, An analytical approach to single node
  delay-coupled reservoir computing\href
  {http://dx.doi.org/10.1007/978-3-642-40728-4_4}
  {\path{doi:10.1007/978-3-642-40728-4_4}}.

\bibitem{SCH13l}
J.~Schumacher, H.~Toutounji, G.~Pipa, {An Analytical Approach to Single Node
  Delay-Coupled Reservoir Computing}, Conference: 23rd International Conference
  on Artificial Neural Networks (2013).
\newblock \href {http://dx.doi.org/10.1007/978-3-642-40728-4_4}
  {\path{doi:10.1007/978-3-642-40728-4_4}}.

\bibitem{PAQ12}
Y.~Paquot, F.~Duport, A.~Smerieri, J.~Dambre, B.~Schrauwen, M.~Haelterman,
  S.~Massar, {Optoelectronic Reservoir Computing}, Sci. Rep. 2~(287).
\newblock \href {http://dx.doi.org/10.1038/srep00287}
  {\path{doi:10.1038/srep00287}}.

\bibitem{Atiya2000}
A.~F. Atiya, A.~G. Parlos, {New Results on Recurrent Network Training: Unifying
  the Algorithms and Accelerating Convergence}, IEEE Trans. Neural Networks
  11~(3).

\bibitem{Weigend1993}
A.~S. {Weigend}, N.~A. {Gershenfeld}, Results of the time series prediction
  competition at the santa fe institute, in: IEEE International Conference on
  Neural Networks, 1993, pp. 1786--1793 vol.3.
\newblock \href {http://dx.doi.org/10.1109/ICNN.1993.298828}
  {\path{doi:10.1109/ICNN.1993.298828}}.

\bibitem{Larger2017}
L.~Larger, A.~Bayl\'on-Fuentes, R.~Martinenghi, V.~S. Udaltsov, Y.~K. Chembo,
  M.~Jacquot,
  \href{https://link.aps.org/doi/10.1103/PhysRevX.7.011015}{High-speed photonic
  reservoir computing using a time-delay-based architecture: Million words per
  second classification}, Phys. Rev. X 7 (2017) 011015.
\newblock \href {http://dx.doi.org/10.1103/PhysRevX.7.011015}
  {\path{doi:10.1103/PhysRevX.7.011015}}.
\newline\urlprefix\url{https://link.aps.org/doi/10.1103/PhysRevX.7.011015}

\bibitem{Haykin1995}
S.~Haykin, {Adaptive filter theory}, 3rd Edition, Prentice Hall, 1995.

\bibitem{DAM12}
J.~Dambre, D.~Verstraeten, B.~Schrauwen, S.~Massar, {Information processing
  capacity of dynamical systems}, Sci. Rep. 2 (2012) 514.
\newblock \href {http://dx.doi.org/10.1038/srep00514}
  {\path{doi:10.1038/srep00514}}.

\end{thebibliography}

\end{document}